\PassOptionsToPackage{unicode}{hyperref}
\PassOptionsToPackage{hyphens}{url}
\PassOptionsToPackage{dvipsnames,svgnames,x11names}{xcolor}
\documentclass[
  english,
  11pt,
  a4paper,
  DIV=11,
  numbers=noendperiod]{scrartcl}
\usepackage{xcolor}
\usepackage[margin=1in]{geometry}
\usepackage{amsmath,amssymb}
\usepackage{iftex}
\ifPDFTeX
  \usepackage[T1]{fontenc}
  \usepackage[utf8]{inputenc}
  \usepackage{textcomp} 
\else 
  \usepackage{unicode-math} 
  \defaultfontfeatures{Scale=MatchLowercase}
  \defaultfontfeatures[\rmfamily]{Ligatures=TeX,Scale=1}
\fi
\usepackage{lmodern}
\ifPDFTeX\else
\fi
\IfFileExists{upquote.sty}{\usepackage{upquote}}{}
\IfFileExists{microtype.sty}{
  \usepackage[]{microtype}
  \UseMicrotypeSet[protrusion]{basicmath} 
}{}
\makeatletter
\@ifundefined{KOMAClassName}{
  \IfFileExists{parskip.sty}{%
    \usepackage{parskip}
  }{
    \setlength{\parindent}{0pt}
    \setlength{\parskip}{6pt plus 2pt minus 1pt}}
}{
  \KOMAoptions{parskip=half}}
\makeatother
\makeatletter
\ifx\paragraph\undefined\else
  \let\oldparagraph\paragraph
  \renewcommand{\paragraph}{
    \@ifstar
      \xxxParagraphStar
      \xxxParagraphNoStar
  }
  \newcommand{\xxxParagraphStar}[1]{\oldparagraph*{#1}\mbox{}}
  \newcommand{\xxxParagraphNoStar}[1]{\oldparagraph{#1}\mbox{}}
\fi
\ifx\subparagraph\undefined\else
  \let\oldsubparagraph\subparagraph
  \renewcommand{\subparagraph}{
    \@ifstar
      \xxxSubParagraphStar
      \xxxSubParagraphNoStar
  }
  \newcommand{\xxxSubParagraphStar}[1]{\oldsubparagraph*{#1}\mbox{}}
  \newcommand{\xxxSubParagraphNoStar}[1]{\oldsubparagraph{#1}\mbox{}}
\fi
\makeatother

\usepackage{longtable,booktabs,array}
\usepackage{calc} 
\usepackage{etoolbox}
\makeatletter
\patchcmd\longtable{\par}{\if@noskipsec\mbox{}\fi\par}{}{}
\makeatother
\IfFileExists{footnotehyper.sty}{\usepackage{footnotehyper}}{\usepackage{footnote}}
\makesavenoteenv{longtable}
\usepackage{graphicx}
\makeatletter
\newsavebox\pandoc@box
\newcommand*\pandocbounded[1]{
  \sbox\pandoc@box{#1}%
  \Gscale@div\@tempa{\textheight}{\dimexpr\ht\pandoc@box+\dp\pandoc@box\relax}%
  \Gscale@div\@tempb{\linewidth}{\wd\pandoc@box}%
  \ifdim\@tempb\p@<\@tempa\p@\let\@tempa\@tempb\fi
  \ifdim\@tempa\p@<\p@\scalebox{\@tempa}{\usebox\pandoc@box}%
  \else\usebox{\pandoc@box}%
  \fi%
}
\def\fps@figure{htbp}
\makeatother

\NewDocumentCommand\citeproctext{}{}

\makeatletter
 \let\@cite@ofmt\@firstofone
 \def\@biblabel#1{}
 \def\@cite#1#2{{#1\if@tempswa , #2\fi}}
\makeatother
\newlength{\cslhangindent}
\newlength{\csllabelwidth}
\newenvironment{CSLReferences}[2] 
 {\begin{list}{}{%
  \setlength{\itemindent}{0pt}
  \setlength{\leftmargin}{0pt}
  \setlength{\parsep}{0pt}
  \ifodd #1
   \setlength{\leftmargin}{\cslhangindent}
   \setlength{\itemindent}{-1\cslhangindent}
  \fi
  \setlength{\itemsep}{#2\baselineskip}}}
 {\end{list}}
\usepackage{calc}

\ifLuaTeX
\usepackage[bidi=basic,shorthands=off]{babel}
\else
\usepackage[bidi=default,shorthands=off]{babel}
\fi
\ifLuaTeX
  \usepackage{selnolig} 
\fi

\providecommand{\tightlist}{%
  \setlength{\itemsep}{0pt}\setlength{\parskip}{0pt}}

\usepackage[T1]{fontenc}
\usepackage{needspace}
\KOMAoption{captions}{tableheading}
\makeatletter
\@ifpackageloaded{caption}{}{\usepackage{caption}}
\AtBeginDocument{%
\ifdefined\contentsname
  \renewcommand*\contentsname{Table of contents}
\else
  \newcommand\contentsname{Table of contents}
\fi
\ifdefined\listfigurename
  \renewcommand*\listfigurename{List of Figures}
\else
  \newcommand\listfigurename{List of Figures}
\fi
\ifdefined\listtablename
  \renewcommand*\listtablename{List of Tables}
\else
  \newcommand\listtablename{List of Tables}
\fi
\ifdefined\figurename
  \renewcommand*\figurename{Figure}
\else
  \newcommand\figurename{Figure}
\fi
\ifdefined\tablename
  \renewcommand*\tablename{Table}
\else
  \newcommand\tablename{Table}
\fi
}
\@ifpackageloaded{float}{}{\usepackage{float}}
\floatstyle{ruled}
\@ifundefined{c@chapter}{\newfloat{codelisting}{h}{lop}}{\newfloat{codelisting}{h}{lop}[chapter]}
\floatname{codelisting}{Listing}

\makeatother
\makeatletter
\@ifpackageloaded{caption}{}{\usepackage{caption}}
\@ifpackageloaded{subcaption}{}{\usepackage{subcaption}}
\makeatother
\usepackage{bookmark}
\IfFileExists{xurl.sty}{\usepackage{xurl}}{} 
\makeatletter
\@ifundefined{xmpquote}{}{}
\makeatother
\hypersetup{
  pdftitle={Inherited Wage Dispersion and Optimal Discretion in a Dual-Rigidity TANK Model},
  pdfauthor={Kenji Miyazaki},
  pdflang={en},
  colorlinks=true,
  linkcolor={blue},
  filecolor={Maroon},
  citecolor={Blue},
  urlcolor={Blue},
  pdfcreator={LaTeX via pandoc}}

\title{Inherited Wage Dispersion and Optimal Discretion in a
Dual-Rigidity TANK Model}
\author{Kenji Miyazaki\footnote{Faculty of Economics, Hosei University.}}
\date{}
\begin{document}
\maketitle
\begin{abstract}
How does an inherited cross-type wage gap enter Markov-perfect
discretionary monetary policy when transfers are passive? In a two-agent
New Keynesian model with sticky prices and type-specific own-lag wage
adjustment, the gap changes implementable allocations and the
second-order welfare loss. A positive lower bound establishes its value
relevance; explicit rank conditions characterize when current price
inflation, wage inflation, and the output gap fail to determine the
implementing nominal rate. An illustrative parameterization satisfies
these conditions, although the additional state explains little
nominal-rate variance after conditioning on all three aggregate
variables. Welfare comparisons with fixed rules are driven mainly by
aggregate wage-inflation stabilization and do not isolate the value of
distributional information. Unrestricted targeted transfers separate
aggregate allocation from the legacy wage-gap transition. In the
CES-consistent distribution block, optimal smoothing eliminates
consumption dispersion, improves on immediate wage-gap elimination, and
coincides under discretion and date-0 commitment.


\end{abstract}

\section{Introduction}\label{sec-introduction}

Discretionary monetary policy in a dual-rigidity economy already
inherits an aggregate real-wage state. This paper asks whether
type-specific wage adjustment adds a distinct distributional state to
that policy problem. In a two-agent economy with passive transfers, past
relative-wage adjustment leaves a cross-type wage gap. The gap affects
feasible allocations, and changing it creates a welfare cost because
each wage adjusts relative to its own lag. The policy authority
therefore inherits a distributional stabilization problem even though it
has made no promises about future policy.\footnote{This private-state
  dependence differs from commitment and timeless-perspective policy,
  where lagged promises or multipliers enter the state. For a
  heterogeneous-agent treatment of those policy concepts, see Dávila and
  Schaab (2023).}

The model combines hand-to-mouth households and savers with sticky
prices and type-specific sticky wages. Labor services and wage setting
are segmented by financial type, so financial status and labor-market
segment coincide by assumption. The central bank chooses the nominal
interest rate, while exogenous transfers are financed contemporaneously
by savers. Aggregate wage rigidity creates the familiar trade-off among
price inflation, wage inflation, and the output gap. Type-specific
own-lag wage adjustment adds a relative-wage state to that trade-off.
The distinction concerns the state variables needed for policy, rather
than the general possibility of state dependence under discretion.

The analytical results separate three questions. First, inherited wage
dispersion changes the feasible set and has a strictly positive value
cost away from the symmetric state. Second, explicit rank conditions
characterize when the same current price inflation, wage inflation, and
output gap are consistent with different implementing nominal rates.
Third, quantitative exercises measure the size of that conditional
policy difference under a specified shock distribution. A state can be
required for an exact policy representation while adding little to its
stationary explanatory power. These claims require different evidence
and are treated separately.

An illustrative parameterization shows both the mechanism and its
quantitative limits. A higher inherited wage gap raises the
discretionary nominal rate when the other primitive states are held
fixed. Current price inflation, wage inflation, and the output gap
nevertheless explain almost all of the nominal-rate variation under the
specified stationary shock covariance. This small increment in explained
rate variance does not measure the welfare value of observing the state.
Optimal discretion also improves on an optimized price-inflation-only
rule, but most of that loss difference comes from aggregate
wage-inflation stabilization after technology shocks. Richer aggregate
rules substantially close the gap and can outperform discretion under
the stationary loss criterion. All fixed-rule comparisons assume
commitment to future feedback coefficients; none measures the marginal
welfare cost of omitting the distributional state under discretion.

An active-transfer benchmark identifies the institutional boundary of
the result. Unrestricted targeted transfers separate aggregate
allocation from the distributional transition, but do not remove the
cost of a legacy wage gap. Immediate post-impact representative-agent
implementation is feasible and is not optimal from a nonzero gap. Under
the CES-consistent objective, the optimal distributional transition
eliminates consumption dispersion and smooths relative wages. The same
transition solves Markov-perfect discretion and date-0 commitment within
this separated block. This equality does not extend the analysis to a
full monetary-fiscal Ramsey problem.

The paper builds on the dual-rigidity TANK framework of Miyazaki (2026).
That companion paper develops the canonical representation and the
post-impact representative-agent implementation, and notes that
implementation alone does not establish welfare optimality. The
contribution here is the positive welfare-cost bound, the conditional
policy-spanning criterion for passive-transfer discretion, and the
formal welfare comparison and optimal smoothing result under active
transfers. The model and its fiscal closure are stated below so that the
policy and welfare arguments can be assessed within this paper.

Section~\ref{sec-related-literature} reviews the related literature.
Section~\ref{sec-model} presents the model and welfare criterion.
Section~\ref{sec-passive-policy} characterizes discretion under passive
transfers. Section~\ref{sec-active-transfers} studies the legacy
wage-gap transition under active transfers and compares discretion with
commitment. Section~\ref{sec-quantitative} presents the numerical
analysis, first under passive transfers and then under active transfers.
Section~\ref{sec-conclusion} concludes. Appendices A--C provide the
welfare derivation, the recursive policy system, and the proofs. The
online appendix contains the full recursive derivation, numerical
methods, and supplementary results.

\section{Related Literature on Optimal Monetary
Policy}\label{sec-related-literature}

This paper sits at the intersection of three literatures: optimal
monetary policy with price and wage rigidities, heterogeneous-agent
monetary transmission, and monetary-fiscal stabilization. Existing work
studies each force in depth. What remains less clear is how an inherited
distributional state enters discretionary monetary policy when wage
adjustment is costly and fiscal transfers are passive.

\subsection{Optimal monetary policy with price and wage rigidities in
representative-agent
models}\label{optimal-monetary-policy-with-price-and-wage-rigidities-in-representative-agent-models}

Representative-agent New Keynesian models with sticky prices and sticky
wages provide the natural benchmark. Erceg et al. (2000) show that
welfare depends on the output gap, price inflation, and wage inflation,
so strict price-inflation targeting is generally not optimal. Blanchard
and Galí (2007) reinforce the broader lesson that labor-market
distortions alter optimal stabilization trade-offs. In that literature,
aggregate wage inflation is a target variable, but there is no
type-specific relative wage to inherit. I ask how the same wage-rigidity
margin changes when wage rigidity is type specific and relative wages
become a cross-type state inherited from the private sector.

\subsection{Optimal monetary policy with heterogeneous
agents}\label{optimal-monetary-policy-with-heterogeneous-agents}

HANK research shows that monetary policy affects aggregate demand not
only through intertemporal substitution, but also through redistribution
across households with different marginal propensities to consume,
income exposures, and balance sheets (Kaplan et al. 2018; Auclert 2019).
Optimal-policy work asks how these channels should enter the policy
problem. Bhandari et al. (2021) study joint monetary-fiscal policy with
heterogeneous agents, incomplete markets, and nominal rigidities, and
show how insurance motives can dominate price-stabilization motives.
Acharya et al. (2023) characterize a target criterion that internalizes
consumption-risk and distributional trade-offs. Dávila and Schaab (2023)
show in a canonical HANK model with wage rigidity that distributional
motives can modify discretion, commitment, and timeless policy. My
commitment calculation is narrower: it applies only to the separated
distribution block after unrestricted active transfers, not to a full
HANK Ramsey comparison. Tractable heterogeneous-agent models also show
that limited asset-market participation can add financial-stability or
liquidity-insurance motives to monetary policy (Nisticò 2016; Bilbiie
and Ragot 2021).

In the passive-transfer baseline, I abstract from balance sheets,
liquidity insurance, and active redistributive policy to isolate whether
an inherited wage-gap state changes discretion when the central bank
controls only the interest rate. This focus is consistent with McKay and
Wolf (2023), who emphasize that monetary policy is a blunt
redistributive instrument. McKay and Wolf (2026) study interest-rate and
fiscal-transfer rules in HANK and find a limited role for monetary
redistribution alongside a larger role for fiscal transfers. Recent HANK
surveys likewise organize redistribution and stabilization as a joint
monetary-fiscal problem (Auclert et al. 2025). Inherited inequality can
change the stabilization trade-off even when monetary policy has limited
redistributive effects.

\subsection{TANK as a tractable bridge between RANK and
HANK}\label{tank-as-a-tractable-bridge-between-rank-and-hank}

TANK and analytical HANK/THANK models provide the tractable bridge used
here. They preserve the aggregate-demand effects of limited asset-market
participation while keeping the state space small (Bilbiie 2008, 2025;
Debortoli and Galí 2025). Broer et al. (2020) show that a tractable
heterogeneous-agent New Keynesian model can have a two-agent reduced
form and that wage rigidity changes monetary transmission. Sticky-wage
TANK models are especially close because wage rigidity links
redistribution to nominal adjustment. Colciago (2011) shows how nominal
wage rigidity changes determinacy and transmission in a rule-of-thumb
consumer model. Ascari et al. (2017) combine limited asset-market
participation with sticky prices and sticky wages to study
monetary-policy design. Recent sticky-wage HANK work also shows that the
labor-supply assumptions embedded in wage setting can change wage and
inflation dynamics (Gerke et al. 2024).\footnote{Ida and Okano (2024)
  study fiscal multipliers in a sticky-wage TANK environment, which is
  complementary to the monetary-policy question studied here.}

Heterogeneous wage contracts are another close comparison. Matsui and
Yoshimi (2013) study monetary-policy rules when wage rigidity differs
across worker groups and emphasize the welfare role of wage-inflation
stabilization. Matsui and Yoshimi (2015) model union and nonunion labor
as directly substitutable inputs within firms. These models motivate
treating wage heterogeneity as a separate modeling choice. Here the
distinctive assumption aligns financial types with labor-market
segments, and own-lag wage adjustment makes their relative wage an
inherited state in the passive-transfer discretionary problem.

The TANK structure makes that inherited state explicit. Miyazaki (2026)
develops the dual-rigidity TANK canonical representation and gives an
active-transfer implementation of the post-impact RANK path. I use the
same canonical environment to characterize passive-transfer discretion
in a finite-dimensional recursive system and to compare post-impact RANK
implementation with the welfare-optimal active-transfer transition. The
passive policy coefficients are solved numerically; the closed-form
smoothing result applies to the separated active-transfer block.

\subsection{Stabilization versus redistribution and the role of fiscal
instruments}\label{stabilization-versus-redistribution-and-the-role-of-fiscal-instruments}

In heterogeneous-agent economies, stabilization and redistribution are
often jointly determined. Auclert (2019) highlights the role of
redistribution in monetary transmission, while Bilbiie et al. (2024)
show that with heterogeneity and nominal rigidities the optimal policy
problem is naturally monetary-fiscal rather than purely monetary. For a
static imperfect-insurance wedge, Bilbiie et al. (2024) show that active
transfers can implement perfect insurance and restore an aggregate RANK
allocation. La'O and Morrison (2025) characterize redistributive
monetary policy with complete markets, type-specific productivity, and
constrained fiscal instruments; state-contingent markups respond to
distributional objectives. Their mechanism differs from the inherited
relative-wage transition studied here. The baseline model and policy
comparisons here retain passive transfers. The active-transfer benchmark
permits unrestricted targeted transfers to isolate the dynamic
distributional transition and to compare discretion with date-0
commitment in the separated distribution block.

Table~\ref{tbl-positioning} compares selected benchmarks. The
contribution concerns a particular private wage state and its role in
discretion: a positive welfare-cost bound, a conditional test of
aggregate policy spanning, and the optimal fiscal transition after
aggregate allocation separates. It does not rely on distributional
motives or wage heterogeneity being absent from the broader HANK and
sticky-wage literatures.

\begingroup
\footnotesize

\begin{longtable}[]{@{}
  >{\raggedright\arraybackslash}p{(\linewidth - 10\tabcolsep) * \real{0.1600}}
  >{\raggedright\arraybackslash}p{(\linewidth - 10\tabcolsep) * \real{0.1500}}
  >{\raggedright\arraybackslash}p{(\linewidth - 10\tabcolsep) * \real{0.1600}}
  >{\raggedright\arraybackslash}p{(\linewidth - 10\tabcolsep) * \real{0.1700}}
  >{\raggedright\arraybackslash}p{(\linewidth - 10\tabcolsep) * \real{0.1400}}
  >{\raggedright\arraybackslash}p{(\linewidth - 10\tabcolsep) * \real{0.2200}}@{}}
\caption{Selected policy benchmarks and the companion
paper.}\label{tbl-positioning}\tabularnewline
\toprule\noalign{}
\begin{minipage}[b]{\linewidth}\raggedright
Margin
\end{minipage} & \begin{minipage}[b]{\linewidth}\raggedright
Dual-rigidity RANK
\end{minipage} & \begin{minipage}[b]{\linewidth}\raggedright
HANK policy
\end{minipage} & \begin{minipage}[b]{\linewidth}\raggedright
Sticky-wage TANK
\end{minipage} & \begin{minipage}[b]{\linewidth}\raggedright
Companion
\end{minipage} & \begin{minipage}[b]{\linewidth}\raggedright
This paper
\end{minipage} \\
\midrule\noalign{}
\endfirsthead
\toprule\noalign{}
\begin{minipage}[b]{\linewidth}\raggedright
Margin
\end{minipage} & \begin{minipage}[b]{\linewidth}\raggedright
Dual-rigidity RANK
\end{minipage} & \begin{minipage}[b]{\linewidth}\raggedright
HANK policy
\end{minipage} & \begin{minipage}[b]{\linewidth}\raggedright
Sticky-wage TANK
\end{minipage} & \begin{minipage}[b]{\linewidth}\raggedright
Companion
\end{minipage} & \begin{minipage}[b]{\linewidth}\raggedright
This paper
\end{minipage} \\
\midrule\noalign{}
\endhead
\bottomrule\noalign{}
\endlastfoot
Household structure & Single agent & Asset and productivity
heterogeneity & Two financial types & Two financial types & Two
financial types \\
Wage setting & Aggregate wage rigidity & Union wage rigidity & Common
wage across financial types & Type-specific own-lag adjustment &
Type-specific own-lag adjustment \\
Policy exercise & Optimal stabilization & Discretion, commitment, and
timeless policy & Transmission, determinacy, and policy rules &
Post-impact RANK path & Passive discretion; active smoothing \\
Inherited wage-gap role & No cross-type wage gap & Distributional policy
motives; no separate H--S wage-gap state & Aggregate stabilization &
Post-impact RANK path & Welfare-cost bound; policy spanning; optimal
fiscal transition \\
\end{longtable}

\emph{Notes:} The first three columns refer respectively to Erceg et al.
(2000), Dávila and Schaab (2023), and Colciago (2011) together with
Ascari et al. (2017). The companion column refers to Miyazaki (2026).
The entries describe these selected benchmarks, not every model in each
literature. Wage-contract heterogeneity is considered separately in the
discussion of Matsui and Yoshimi (2013) and Matsui and Yoshimi (2015).

\endgroup

\section{Model and Welfare}\label{sec-model}

I use a dual-rigidity TANK economy in which nominal rigidities link
aggregate stabilization to redistribution. Hand-to-mouth households
spend current disposable income. Savers trade bonds and receive profits.
Changes in wages, transfers, and markups therefore move purchasing power
across household types. Price rigidity makes aggregate inflation costly,
while own-lag wage rigidity makes relative wages adjust gradually. This
section states the household, firm, and fiscal arrangements, derives the
canonical equilibrium system, and defines the second-order welfare loss
and recursive discretionary problem.

The key inherited object is the cross-type wage gap. The wage-setting
structure turns that gap into a persistent distributional state,
summarized below by the net distributional wedge \(\omega_t\) and the
law of motion for \(\sigma_t^w\). Transfers are passive, so fiscal
policy does not remove the inherited state before monetary policy acts.
The policy question is how a discretionary central bank should set
\(i_t\) when inherited wage inequality already affects aggregate demand
and welfare.

The underlying dual-rigidity TANK environment is common to Miyazaki
(2026), which develops the canonical representation. The household,
firm, and fiscal arrangements supporting that representation are stated
here. The new object here is not the canonical representation itself,
but the recursive discretionary monetary-policy problem generated by
that representation under passive transfers.
\hyperref[sec-appendix-a]{Appendix A} derives the second-order
welfare-based objective, \hyperref[sec-appendix-b]{Appendix B} reports
the reduced recursive policy system, and the online appendix gives the
full first-order-condition algebra.

\subsection{Economic environment}\label{sec-model-foundations}

The economy contains a unit continuum of households. Households in
\(\mathcal H=[0,\lambda)\) are hand-to-mouth, and those in
\(\mathcal S=[\lambda,1]\) are savers. Each household supplies a
differentiated labor service and sets its own nominal wage. Identical
initial conditions and common shocks within each financial type imply
within-type symmetry, with consumption, employment, and real wages
denoted by \((C_t^j,N_t^j,W_t^j)\) for \(j\in\{H,S\}\). This
deliberately aligns financial types with labor-market segments. If the
two financial types instead supplied the same mixture of labor services,
a relative wage between those services would not map into the H--S
earnings gap in the same way. The results below are conditional on the
maintained alignment.

For the level variables in this model description, omitting the time
subscript denotes the steady-state value. Thus \(C_t\) is aggregate
consumption at time \(t\), and \(C\) is its steady-state level.
Superscript \(N\) denotes nominal variables, and \(E_t\) denotes
expectation conditional on information available at time \(t\).

\subsubsection{Households and labor
demand}\label{households-and-labor-demand}

All households maximize \[
E_0\sum_{t=0}^{\infty}\beta^t
\left[\frac{C_t(h)^{1-\gamma}}{1-\gamma}
-\frac{N_t(h)^{1+\varphi}}{1+\varphi}\right],
\] Here \(\beta\in(0,1)\) is the discount factor, \(\gamma>0\) is the
inverse intertemporal elasticity of substitution, and \(\varphi\ge0\) is
the inverse Frisch elasticity of labor supply; logarithmic consumption
utility is understood when \(\gamma=1\). A competitive labor packer
combines differentiated services with elasticity \(\psi_w>1\). Its
demand schedule and the corresponding type-symmetric quantity and
nominal-wage indices are \[
\begin{aligned}
N_t(h)&=N_t\left(\frac{W_t^N(h)}{W_t^N}\right)^{-\psi_w},\\
N_t&=\left[\lambda(N_t^H)^{\rho_w}
+(1-\lambda)(N_t^S)^{\rho_w}\right]^{1/\rho_w},
\qquad \rho_w=1-\frac1{\psi_w},\\
W_t^N&=\left[\lambda(W_t^{N,H})^{1-\psi_w}
+(1-\lambda)(W_t^{N,S})^{1-\psi_w}\right]^{1/(1-\psi_w)}.
\end{aligned}
\] Let \(P_t\) denote the aggregate price index. Real wages satisfy
\(W_t(h)=W_t^N(h)/P_t\) and \(W_t=W_t^N/P_t\). Changing an individual
nominal wage uses final goods in the amount \[
\mathcal A_t^w(h)=\frac{\eta_w}{2}
\left(\frac{W_t^N(h)}{W_{t-1}^N(h)}-1\right)^2Y_t,
\qquad \eta_w>0.
\] The household pays this cost. The denominator is its own previous
wage, so today's wage choice also affects tomorrow's adjustment cost.

Let \(T_t^H\) and \(T_t^S\) be per-household lump-sum fiscal
liabilities, let \(D_t\) denote total firm dividends, and let \(R_t^N\)
be the gross nominal return on a bond purchased at \(t\). Nominal bond
holdings are denoted by \(B_t^N(h)\). The household budgets are \[
\begin{aligned}
C_t(h)+\mathcal A_t^w(h)+T_t^H
&=(1+\tau_w)W_t(h)N_t(h), &&h\in\mathcal H,\\
C_t(h)+\mathcal A_t^w(h)+\frac{B_t^N(h)}{P_t}+T_t^S
&=(1+\tau_w)W_t(h)N_t(h)\\
&\quad+\frac{R_{t-1}^N B_{t-1}^N(h)}{P_t}
+\frac{D_t}{1-\lambda}, &&h\in\mathcal S.
\end{aligned}
\] Savers own firms equally and trade one-period nominal bonds in zero
net supply, subject to the usual no-Ponzi restriction and transversality
condition. Hand-to-mouth households cannot trade bonds. Each household
takes aggregate variables and its type's lump-sum liability as given
when choosing its wage and consumption. The saver Euler equation is \[
1=\beta E_t\left[
R_t^N\left(\frac{C_{t+1}^S}{C_t^S}\right)^{-\gamma}
\frac{P_t}{P_{t+1}}\right].
\]

\subsubsection{Firms and the fiscal
accounts}\label{firms-and-the-fiscal-accounts}

A unit continuum of intermediate-good firms uses \(Y_t(v)=A_tN_t(v)\),
where \(A_t=\exp(a_t)\). Competitive final-good aggregation gives demand
\(Y_t(v)=Y_t[P_t(v)/P_t]^{-\psi_p}\) with \(\psi_p>1\). Firm \(v\)
maximizes the expected present value of its real dividends using the
savers' discount factor, \[
E_0\sum_{t=0}^{\infty}\beta^t
\left(\frac{C_t^S}{C_0^S}\right)^{-\gamma}D_t(v),
\] subject to demand, production, and \[
\begin{aligned}
D_t(v)&=(1+\tau_p)\frac{P_t(v)}{P_t}Y_t(v)
-W_tN_t(v)-\mathcal A_t^p(v)-T_t^M,\\
\mathcal A_t^p(v)&=\frac{\eta_p}{2}
\left(\frac{P_t(v)}{P_{t-1}(v)}-1\right)^2Y_t,
\qquad \eta_p>0.
\end{aligned}
\] The price-adjustment cost therefore reduces dividends and uses final
goods. Firms take the uniform lump-sum levy \(T_t^M\) and all aggregate
variables as given. Common initial prices and aggregate shocks imply a
symmetric-price equilibrium, so \(Y_t=A_tN_t\) and
\(D_t=\int_0^1D_t(v)\,dv\).

The fiscal closure finances the markup-correcting subsidies within their
respective tax bases. Define average pre-subsidy labor earnings within
each financial type by \[
\mathcal E_t^H=\frac1\lambda\int_{\mathcal H}W_t(h)N_t(h)\,dh,
\qquad
\mathcal E_t^S=\frac1{1-\lambda}\int_{\mathcal S}W_t(h)N_t(h)\,dh.
\] For the heterogeneous economy, \(0<\lambda<1\), set \[
\begin{aligned}
\tau_w&=\frac1{\psi_w-1},&
\tau_p&=\frac1{\psi_p-1},\\
T_t^H&=\tau_w\mathcal E_t^H-\mathcal Z_t,&
T_t^S&=\tau_w\mathcal E_t^S+\frac{\lambda}{1-\lambda}\mathcal Z_t,
\qquad T_t^M=\tau_pY_t.
\end{aligned}
\] Here \(\mathcal Z_t=Yz_t\) is the transfer received by each
hand-to-mouth household, with zero steady-state transfer and
\(z_t=\rho_z z_{t-1}+e_t^z\). Savers finance that transfer
contemporaneously. In the active-transfer benchmark, replace
\(\mathcal Z_t\) by \(Y(z_t+f_t)\); the additional transfer obeys the
same budget accounting.

The subsidy rates are constant, while their expenditures and financing
levies vary with the corresponding aggregate tax bases. Each liability
is common to a continuum of households or firms and is unaffected by an
individual agent's choice. Thus the levies are lump sum at the
individual margin even though they vary in equilibrium. This distinction
is essential: applying a levy to an individual's own earnings before
taking its wage derivative would cancel the marginal subsidy and produce
a different model. The present closure makes explicit the fiscal
incidence underlying the canonical representation in Miyazaki (2026).

The government balances its budget every period: \[
\lambda T_t^H+(1-\lambda)T_t^S+T_t^M
=\tau_w\left[\lambda\mathcal E_t^H+(1-\lambda)\mathcal E_t^S\right]
+\tau_pY_t.
\] There is no public debt or government consumption. Within-type
symmetry makes the wage subsidy and its financing levy cancel in each
type's average budget, while the price subsidy and firm levy cancel in
total dividends. Consequently, \[
\begin{aligned}
C_t^H&=W_t^HN_t^H+\mathcal Z_t-\mathcal A_t^{w,H},\\
D_t&=Y_t-W_tN_t-\mathcal A_t^p,\\
Y_t&=C_t+\mathcal A_t^p
+\lambda\mathcal A_t^{w,H}+(1-\lambda)\mathcal A_t^{w,S},
\qquad C_t=\lambda C_t^H+(1-\lambda)C_t^S.
\end{aligned}
\] Here \(\mathcal A_t^p=\int_0^1\mathcal A_t^p(v)\,dv\) is total
price-adjustment cost, and \(\mathcal A_t^{w,j}\) is the wage-adjustment
cost of a type-\(j\) household under within-type symmetry. These
cancellations are equilibrium accounting identities. The subsidies
remain in the individual price- and wage-setting first-order conditions.

\subsubsection{Wage setting, price setting, and the reference
allocation}\label{wage-setting-price-setting-and-the-reference-allocation}

Define \(\Pi_t^j=W_t^{N,j}/W_{t-1}^{N,j}\), \(\Pi_t^p=P_t/P_{t-1}\), and
\(MRS_t^j=(C_t^j)^\gamma(N_t^j)^\varphi\). With the specified subsidies,
the type-symmetric wage-setting condition is \[
\begin{aligned}
\eta_w\Pi_t^j(\Pi_t^j-1)
={}&\psi_w\frac{W_t^jN_t^j}{Y_t}
\left(\frac{MRS_t^j}{W_t^j}-1\right)\\
&+\beta\eta_w E_t\left[
\left(\frac{C_{t+1}^j}{C_t^j}\right)^{-\gamma}
\frac{Y_{t+1}}{Y_t}\Pi_{t+1}^j(\Pi_{t+1}^j-1)\right],
\qquad j\in\{H,S\}.
\end{aligned}
\] Hand-to-mouth status does not remove the continuation term: an
inherited nominal wage affects future adjustment costs even without
saving. The symmetric price-setting condition is \[
\begin{aligned}
\eta_p\Pi_t^p(\Pi_t^p-1)
={}&\psi_p\left(\frac{W_t}{A_t}-1\right)\\
&+\beta\eta_p E_t\left[
\left(\frac{C_{t+1}^S}{C_t^S}\right)^{-\gamma}
\frac{Y_{t+1}}{Y_t}\Pi_{t+1}^p(\Pi_{t+1}^p-1)\right].
\end{aligned}
\] The subsidy identities are \((1+\tau_w)(1-1/\psi_w)=1\) for wages and
\((1+\tau_p)(1-1/\psi_p)=1\) for prices. At the zero-inflation steady
state with \(A=1\) and \(\mathcal Z=0\), they imply \(W=MRS=1\). The
budgets, resource constraint, and common preferences then give \[
C^H=C^S=C=N^H=N^S=N=Y=1,
\qquad D=0,
\qquad R^N=\beta^{-1}.
\] This allocation is symmetric and efficient. In particular,
\(N^{1+\varphi}/C^{1-\gamma}=1\), which justifies the unit coefficient
on the labor term in the normalized welfare expansion in Appendix A.
Symmetry alone would not justify that coefficient.

The subsidies remove the steady-state markup distortions; the Rotemberg
terms remain real adjustment costs away from steady state. Lower-case
quantity and wage variables denote log deviations from this steady
state, and \(\pi_t^j=\log\Pi_t^j\) and \(\pi_t^p=\log\Pi_t^p\).
Linearizing the optimality conditions gives \[
\begin{aligned}
\pi_t^j-\beta E_t\pi_{t+1}^j
&=\frac{\psi_w}{\eta_w}
\left(\gamma c_t^j+\varphi n_t^j-w_t^j\right),\\
\pi_t^p-\beta E_t\pi_{t+1}^p
&=\frac{\psi_p}{\eta_p}(w_t-a_t),\\
c_t^H&=w_t^H+n_t^H+z_t.
\end{aligned}
\] Thus the same fiscal closure delivers the household budget
coefficients and Phillips-curve slopes used below. The second-order
resource constraint retains the adjustment costs with coefficients
\(\eta_p\) and \(\eta_w\), while CES labor aggregation supplies the
additional \(1/\psi_w\) labor-dispersion weight. The loss used in the
paper is exact within this maintained second-order approximation around
the efficient symmetric steady state.

\subsection{Canonical representation}\label{sec-canonical}

The canonical representation keeps aggregate variables and cross-type
dispersion separate. Four objects drive the mechanism: the TANK IS
curve, which transmits cross-type consumption dispersion into aggregate
demand; the type-specific wage-dispersion law of motion, which makes
\(\sigma_{t-1}^w\) an inherited state; the welfare loss, which penalizes
abrupt changes in the wage gap; and the IS implementability condition,
which recovers the interest rate from the chosen allocation. This
subsection defines the first two objects and the surrounding equilibrium
system. The next subsection states the loss function and the recursive
policy problem.

The economics can be read from these four objects before following the
algebra. A transfer or technology shock moves the net distributional
wedge. The wedge changes consumption dispersion and the output gap
through the TANK IS curve. Own-lag wage setting makes the wage gap
persistent. The policymaker then trades off aggregate stabilization
against the welfare cost of moving relative wages too quickly.

At first order, for any \(q \in \{c,w,n\}\), define \[
q_t = \lambda q_t^H + (1-\lambda)q_t^S,
\qquad
\sigma_t^q = q_t^S - q_t^H.
\] The distributional block starts with a net wedge that links aggregate
dynamics to redistribution: \[
\omega_t = a_t - w_t - z_t.
\] The wedge collects the forces that move relative disposable income
before household consumption is allocated across types. A higher
technology level raises firm income relative to wage income, a higher
aggregate real wage shifts income toward workers, and a positive
transfer shock shifts resources toward hand-to-mouth households. Under
the sign convention used here, \(\sigma_t^c\) is saver consumption minus
hand-to-mouth consumption, so a positive transfer shock lowers
\(\omega_t\). The transfer shock is therefore redistribution-sensitive
rather than an aggregate supply disturbance. \footnote{This reduced-form
  wedge is closely related to the transfer wedges that appear in
  tractable TANK models, but here it remains coupled to an endogenous
  wage-dispersion state generated by nominal wage rigidity.} The
associated static dispersion block is \[
\sigma_t^c = (1-\psi_w)\sigma_t^w + \frac{\omega_t}{1-\lambda},
\] \[
\sigma_t^n = -\psi_w \sigma_t^w.
\] The aggregate block defines the output gap, the real interest-rate
gap, and the IS curve. Let \(y_t\) denote log output. The efficient
technology-driven reference allocation and the output gap are \[
y_t^f = \frac{1+\varphi}{\gamma+\varphi}a_t,
\qquad
x_t \equiv y_t-y_t^f.
\] The natural real interest rate \(r_t^f\) is the exogenous
natural-rate process associated with \(y_t^f\). With the log-deviation
normalization used here, \[
r_t^f \equiv \gamma E_t\Delta y_{t+1}^f,
\] Here \(\Delta y_{t+1}^f=y_{t+1}^f-y_t^f\). Let \(R_t^N\) denote the
gross quarterly nominal interest rate and define
\(i_t\equiv\log(R_t^N/R^N)\), where \(R^N=1/\beta\) at the
zero-inflation steady state. The ex ante real-rate log deviation is
\begin{equation}\protect\phantomsection\label{eq-real-nominal-rate}{
r_t\equiv i_t-E_t\pi_{t+1}^p.
}\end{equation} Interest rates and inflation are quarterly log
deviations from their steady-state values; figures do not annualize
them. Thus \(r_t-r_t^f\) is the real rate gap. Redistribution enters
aggregate demand through the TANK IS curve, \[
x_t = E_t x_{t+1} - \frac{1}{\gamma}(i_t-E_t\pi_{t+1}^p-r_t^f) - \lambda(\sigma_t^c-E_t\sigma_{t+1}^c).
\] Two nominal-rigidity equations govern price inflation and the
aggregate real wage. The price Phillips curve is \[
\pi_t^p = \beta E_t \pi_{t+1}^p + \kappa_p(w_t-a_t).
\] The aggregate real wage equation is \[
w_t = \frac{1}{\Theta_w}\left(w_{t-1}+\beta E_t w_{t+1}+\kappa_x x_t + \kappa_a a_t\right).
\] The inherited distributional state evolves according to \[
\sigma_t^w = \frac{1}{\Theta_\sigma}\left(\sigma_{t-1}^w + \beta E_t \sigma_{t+1}^w + \frac{\gamma\kappa_w}{1-\lambda}\omega_t\right).
\] Wage inflation links the price block and the aggregate real wage: \[
\pi_t^w = \pi_t^p + w_t - w_{t-1}.
\] The corresponding cross-type wage-inflation dispersion is \[
\pi_t^S - \pi_t^H = \sigma_t^w - \sigma_{t-1}^w.
\] The exogenous shock block closes the canonical system: \[
a_t = \rho_a a_{t-1} + e_t^a,
\qquad
z_t = \rho_z z_{t-1} + e_t^z.
\] Here \(\rho_a\) and \(\rho_z\) are the persistence parameters of the
technology and transfer shocks, and \(e_t^a\) and \(e_t^z\) are their
innovations. Given the natural-output definition above, the technology
process implies \[
r_t^f
= \gamma\frac{1+\varphi}{\gamma+\varphi}(\rho_a-1)a_t.
\] The remaining parameters collect structural primitives into the
slopes used above.\footnote{The slope definitions are \[
  \begin{aligned}
  \kappa_p &= \frac{\psi_p}{\eta_p},
  \qquad
  \kappa_w = \frac{\psi_w}{\eta_w},
  \qquad
  \kappa_x = \kappa_w(\gamma+\varphi),\\
  \kappa_a &= \kappa_p+\kappa_w,
  \qquad
  \Theta_w = 1+\beta+\kappa_a,\\
  \Theta_\sigma &= 1+\beta+\kappa_w\left[(1-\gamma)+\psi_w(\gamma+\varphi)\right].
  \end{aligned}
  \]} The important feature is not the algebraic form of these composite
coefficients. It is the state dependence in the wage-dispersion
equation: \(\sigma_t^w\) is forward looking and also depends on
\(\sigma_{t-1}^w\). Inherited wage inequality is therefore an endogenous
private-sector state variable.

\subsection{Welfare and the discretionary policy
problem}\label{sec-welfare}

The second-order welfare-based loss function turns inherited inequality
into a current stabilization concern. Cross-sectional wage dispersion by
itself is not the only cost. The loss also contains cross-type
wage-inflation dispersion, which is the change in the wage gap. A
central bank that inherits a large wage gap therefore faces a trade-off:
leaving the gap in place affects consumption dispersion and aggregate
demand, while moving it quickly creates costly relative wage inflation.

Throughout the paper, \(\ell_t\) denotes the quadratic period loss
obtained from a second-order approximation to household welfare around
the efficient symmetric steady state, and \(\mathcal L\) denotes its
expected discounted sum. This objective combines aggregate stabilization
costs with cross-type dispersion costs. The aggregate terms penalize the
output gap, price inflation, and aggregate wage inflation. The
dispersion terms penalize cross-type consumption dispersion, labor
dispersion, and the wage-gap adjustment cost: \[
\begin{aligned}
\mathcal L = E_0 \sum_{t=0}^{\infty}\beta^t \bigg[
&\frac{1}{2}\left\{(\gamma+\varphi)x_t^2
+ \eta_p(\pi_t^p)^2 + \eta_w(\pi_t^w)^2\right\} \\
&+ \frac{\lambda(1-\lambda)}{2}
\left\{\gamma(\sigma_t^c)^2 + \left(\varphi+\frac{1}{\psi_w}\right)(\sigma_t^n)^2
+ \eta_w(\sigma_t^w-\sigma_{t-1}^w)^2\right\}
\bigg].
\end{aligned}
\] Under the maintained own-lag wage-inflation specification, the
welfare-based wage-adjustment term penalizes cross-type wage-inflation
dispersion. Since that dispersion equals the change in the cross-type
wage gap, the policy problem contains a direct motive to smooth the
wage-gap state. Inherited wage dispersion therefore enters the recursive
problem through both the law of motion for \(\sigma_t^w\) and the
current-period objective.

The labor-dispersion coefficient includes the curvature of the CES labor
aggregator. Using \(\sigma_t^n=-\psi_w\sigma_t^w\), its contribution in
wage-gap units is \((\varphi\psi_w^2+\psi_w)(\sigma_t^w)^2\).

Because policy is discretionary, the problem is recursive. A variable
belongs in the state vector only if it is payoff relevant for future
choices. The lagged wage gap satisfies that test for economic reasons,
not just algebraic ones: it affects current consumption dispersion
through the distributional block and current welfare through relative
wage inflation. The natural state vector is \[
s_t = (w_{t-1},\sigma_{t-1}^w,a_t,z_t)^\top.
\] I use a Markov-perfect discretionary equilibrium. At the beginning of
period \(t\), the private wage and shock states \(s_t\) are inherited.
The central bank chooses a current implementable allocation as a
function of \(s_t\), taking the same future policy functions as given.
Private expectations are formed using those future policy functions and
the exogenous shock laws. After the allocation is chosen, the TANK IS
curve uniquely recovers the nominal interest rate that implements it,
conditional on the current and expected future allocation. The inversion
uses the linear IS equation with \(\gamma>0\) and abstracts from the
zero lower bound or other instrument constraints.

Let \(u_t\) collect the current allocation variables chosen in the
primal representation, including \(x_t\), \(\pi_t^p\), \(\pi_t^w\),
\(w_t\), \(\sigma_t^w\), \(\sigma_t^c\), \(\sigma_t^n\), and
\(\omega_t\). The Bellman problem is \[
V(s_t)=\min_{u_t} \left\{\ell_t + \beta E_t[V(s_{t+1})]\right\}
\] subject to the canonical constraints in Section~\ref{sec-canonical}.
Equivalently, the central bank chooses the nominal interest rate \(i_t\)
and the implied feasible allocation. \hyperref[sec-appendix-b]{Appendix
B} writes the recursive problem in primal form.

The primal formulation is a notational device, not an additional policy
instrument. The constraints pin down the implementable allocations, and
the TANK IS curve recovers the unique nominal interest rate consistent
with the chosen allocation. This is why the Bellman problem can be
written over allocation variables even though the central bank controls
only \(i_t\).

The objective penalizes both the wage-gap level, through labor
dispersion, and its change, through relative wage inflation. These
distinct costs make inherited dispersion relevant even when consumption
dispersion can be controlled. The following results establish its role
in implementability and welfare, then state conditions under which
current aggregate outcomes do not determine the implementing nominal
rate.

\section{Discretion under Passive Transfers}\label{sec-passive-policy}

The passive-transfer results distinguish the presence of an inherited
distributional state from its implications for a particular policy
instrument. All results use the local linear equilibrium and the
second-order welfare criterion around the efficient symmetric steady
state. Appendix B gives the recursive policy system, and Appendix C
separates the analytical conditions from their quantitative
verification.

\subsection{Passive transfers and the additional distributional
state}\label{sec-passive-results}

The passive-transfer problem inherits two private wage states. The
aggregate real wage is already a state in a representative-agent economy
with sticky prices and wages. Type-specific own-lag wage adjustment adds
the cross-type wage gap. The results below distinguish the economic role
of that additional state from the information needed to recover the
optimal policy rate from current aggregate outcomes.

The analysis uses the linear equilibrium and second-order welfare loss
around the efficient symmetric steady state. Assume an interior,
time-invariant discretionary equilibrium with finite discounted loss,
\(0<\lambda<1\), \(\eta_w>0\), and
\(W_n\equiv\varphi\psi_w^2+\psi_w>0\). The full recursive target
condition is (\ref{eq-b-target}). The propositions below characterize
the role of its inherited state and the information needed to recover
the implementing policy rate.

\subsubsection{An additional state in implementability and
welfare}\label{an-additional-state-in-implementability-and-welfare}

An inherited wage gap limits how much relative wages can adjust today
and makes that adjustment costly. Leaving the gap in place creates labor
dispersion; closing it immediately creates cross-type wage-inflation
dispersion. These two costs establish that the inherited gap is payoff
relevant even before considering its effects on aggregate stabilization.

\textbf{Proposition 1 (A positive welfare cost of inherited wage
dispersion).} In the passive-transfer economy, the recursive state is \[
s_t=(w_{t-1},\sigma_{t-1}^w,a_t,z_t)^\top.
\] Holding the other three components fixed, different inherited wage
gaps imply different feasible sets for the full current allocation,
conditional on the same continuation policy functions. The gap is also
payoff relevant. Let \(e_\sigma=(0,1,0,0)^\top\), and let \(V^0(s)\)
denote discounted equilibrium loss from initial state \(s\) when all
subsequent innovations are zero. For a pure legacy gap \(s=d e_\sigma\),
\begin{equation}\protect\phantomsection\label{eq-passive-legacy-value-bound}{
V^0(d e_\sigma)-V^0(0)
\ge
\frac{\lambda(1-\lambda)}{2}
\frac{W_n\eta_w}{W_n+\eta_w}d^2
>0
\qquad\text{if }d\ne0.
}\end{equation}

The proposition establishes an additional state for the allocation and
welfare problem. It does not, by itself, imply that every policy
instrument responds to the gap at every parameter value. In particular,
the value of the inherited state and its coefficient in the implementing
nominal-rate rule are different objects. The lower bound also applies
when transfers are active: fiscal instruments can separate aggregate
allocation from the distributional transition, but cannot eliminate both
the labor-dispersion and wage-adjustment costs of a legacy gap.

\subsubsection{Current aggregate outcomes and the implementing nominal
rate}\label{current-aggregate-outcomes-and-the-implementing-nominal-rate}

To assess what current aggregate outcomes reveal about policy, write a
linear discretionary solution as \[
E_t s_{t+1}=A s_t,\qquad
\pi_t^p=P s_t,\qquad
\pi_t^w=K s_t,\qquad
x_t=X s_t,\qquad
r_t=R s_t.
\] Here \(A\) is the conditional state-transition matrix, \(P,K,X,R\)
are coefficient rows, and \(r_t\) is the ex ante real rate. The nominal
rate satisfies \(i_t=r_t+E_t\pi_{t+1}^p\), so its coefficient row is
\begin{equation}\protect\phantomsection\label{eq-passive-nominal-rate-row}{
i_t=I s_t,\qquad I\equiv R+PA.
}\end{equation} Define the map from the state to current aggregate
outcomes by
\begin{equation}\protect\phantomsection\label{eq-passive-aggregate-observation-map}{
H\equiv
\begin{bmatrix}
P\\K\\X
\end{bmatrix}.
}\end{equation} The following condition determines whether these three
outcomes are sufficient to recover the implementing nominal rate.

\textbf{Proposition 2 (Current aggregate outcomes need not determine the
nominal rate).} Consider a linear discretionary equilibrium defined on a
neighborhood of admissible initial states in \(\mathbb R^4\). Suppose
\begin{equation}\protect\phantomsection\label{eq-passive-policy-rank-conditions}{
\operatorname{rank}(H)=3,
\qquad
v\in\ker H,
\qquad
e_\sigma^\top v\ne0,
\qquad
Iv\ne0.
}\end{equation} Then there are arbitrarily close states with identical
current price inflation, wage inflation, and output gaps, but different
inherited wage gaps and different implementing nominal rates.
Consequently, no function of current \((\pi_t^p,\pi_t^w,x_t)\) alone can
recover the nominal-rate policy function throughout that neighborhood.

These are conditions on a solved equilibrium, rather than a claim that
every admissible calibration satisfies them. Online Appendix B.3
evaluates the conditions for the benchmark solution and constructs the
corresponding state comparison. It also reports the direct nominal-rate
sensitivity \(I e_\sigma\), which holds the other primitive states
fixed. This direct sensitivity differs from \(Iv\), which holds the
three current aggregate outcomes fixed by moving several primitive
states together.

The rank condition clarifies the interpretation of policy-function
projections. If \(\operatorname{rank}(H)=3\) and
\(e_\sigma^\top v\ne0\), adding the inherited gap to the three aggregate
outcomes recovers all four state coordinates. Any other fourth variable
that provides independent state information can do the same. An exact
fit therefore establishes linear spanning; it does not measure the
unique importance of wage dispersion. Projection fit also depends on the
shock processes and their stationary covariance. The benchmark
projections quantify approximation under those processes, while
Proposition 1 establishes the economic role of the inherited gap
independently of the projection exercise.

\subsubsection{Aggregate stabilization and shock
transmission}\label{aggregate-stabilization-and-shock-transmission}

\textbf{Remark (Price stabilization and the distributional margin).}
Price inflation is only one component of the welfare criterion.
Aggregate wage inflation and the output gap already enter the
dual-rigidity stabilization problem, and the present model adds
consumption dispersion, labor dispersion, and changes in the cross-type
wage gap. Setting current price inflation to zero does not remove these
terms from the objective or the recursive target condition. Whether
strict price stabilization satisfies that condition at a particular
state depends on the full equilibrium trade-off. The distributional
contribution is therefore assessed through the state results above and
the quantitative exercises, rather than by treating a price-only
stabilization comparison as a measure of the additional channel.

\textbf{Remark (Technology and transfer shocks).} Technology enters the
price Phillips curve, the aggregate real-wage equation, the natural
rate, and the net distributional wedge. The transfer shock enters the
wedge directly. These different entry points distinguish the
disturbances economically, but their equilibrium effects also include
changes in continuation expectations. The shock-specific impulse
responses and welfare calculations report the resulting differences in
the benchmark solution; distinct direct entry points alone do not
establish that every policy coefficient must differ across the two
shocks.

\subsubsection{Limiting cases}\label{limiting-cases}

When \(\lambda=0\), the representative-agent economy retains the
aggregate real-wage state but has no cross-type distributional
stabilization problem. If wages are sticky only at the aggregate level,
the aggregate wage-inflation margin remains while the inherited
cross-type wage gap is absent. Flexible wages remove the own-lag
wage-adjustment state, although a static TANK consumption wedge may
remain. Finally, unrestricted targeted transfers allow aggregate
allocation to separate from the inherited distributional transition. The
active-transfer benchmark below shows why this separation leaves a
welfare-relevant legacy gap and a motive to smooth its adjustment.

\section{Active Transfers and the Legacy Wage
Gap}\label{sec-active-transfers}

Unrestricted targeted transfers restore aggregate allocation separation,
but they do not make an inherited wage gap costless. The benchmark
builds on the active-transfer implementation in Miyazaki (2026), which
studies the same dual-rigidity TANK environment and constructs the
post-impact RANK-implementation path. I use that setup for a narrower
purpose: to ask what remains of the inherited wage-dispersion problem
when fiscal policy can directly choose the distributional wedge. This
benchmark extends the passive-transfer model by adding a targeted fiscal
instrument.

\subsection{Active-transfer benchmark}\label{sec-active-benchmark}

Let \(z_t\) continue to denote the exogenous transfer shock in the
passive-transfer model. Introduce a new targeted policy transfer \(f_t\)
to hand-to-mouth households. The active distributional wedge is \[
\omega_t^A=a_t-w_t-z_t-f_t.
\] The transfer has no aggregate resource cost, no distortionary tax
wedge, no public debt, no adjustment cost, and no bound. The central
bank and fiscal authority are treated as a single cooperating policy
authority in this benchmark. Setting \(f_t=0\) returns to the
passive-transfer model.

The period loss is unchanged. It separates into aggregate and
distributional components, \[
\ell_t=\ell_t^{agg}+\ell_t^{dist},
\] where \[
\ell_t^{agg}
=
\frac12\left[(\gamma+\varphi)x_t^2
+\eta_p(\pi_t^p)^2
+\eta_w(\pi_t^w)^2\right],
\] and \[
\ell_t^{dist}
=
\frac{\lambda(1-\lambda)}2
\left[
\gamma(\sigma_t^c)^2
+\left(\varphi+\frac{1}{\psi_w}\right)(\sigma_t^n)^2
+\eta_w(\sigma_t^w-\sigma_{t-1}^w)^2
\right].
\] The distribution block becomes \[
\sigma_t^w
=
\frac{1}{\Theta_\sigma}
\left[
\sigma_{t-1}^w+\beta E_t\sigma_{t+1}^w
+\frac{\gamma\kappa_w}{1-\lambda}\omega_t^A
\right],
\] \[
\sigma_t^c=(1-\psi_w)\sigma_t^w+
\frac{\omega_t^A}{1-\lambda},
\qquad
\sigma_t^n=-\psi_w\sigma_t^w.
\] After the policy authority chooses the aggregate allocation and the
active distributional wedge, the instruments are recovered from \[
f_t^*=a_t-w_t^*-z_t-\omega_t^{A,*},
\] and \[
i_t^*=E_t\pi_{t+1}^{p,*}+r_t^f+\gamma(E_tx_{t+1}^*-x_t^*)
-\gamma\lambda(\sigma_t^{c,*}-E_t\sigma_{t+1}^{c,*}).
\] The IS equation is therefore an implementability condition. The
instruments remain coupled in implementation even though the allocation
problem separates.

\subsection{Dynamic allocation separation}\label{sec-dynamic-separation}

The active-transfer benchmark gives an aggregate separation result close
in spirit to Bilbiie et al. (2024). Their separation proposition applies
to a static imperfect-insurance wedge. Here, type-specific own-lag wage
rigidity leaves a dynamic distributional transition even after aggregate
allocation separation is restored.

\textbf{Proposition 3 (Dynamic allocation separation under active
transfers).} Under unrestricted targeted transfers, the joint
Markov-perfect allocation problem separates into a dual-rigidity RANK
aggregate stabilization problem and an autonomous distributional
transition problem indexed by inherited wage dispersion.

The aggregate constraints depend only on \((x_t,\pi_t^p,\pi_t^w,w_t)\)
and the aggregate state \((w_{t-1},a_t)\). The distributional
constraints depend only on
\((\sigma_t^w,\sigma_t^c,\sigma_t^n,\omega_t^A)\) and
\(\sigma_{t-1}^w\). Up to terms independent of policy, the value
function can be written as \[
V^A(w_{t-1},\sigma_{t-1}^w,a_t,z_t)
=V^{agg}(w_{t-1},a_t)+V^{dist}(\sigma_{t-1}^w).
\] The aggregate allocation coincides with the dual-rigidity RANK
discretionary allocation. The passive transfer shock \(z_t\) drops out
of the allocation problem because \(f_t\) can offset it in
\(\omega_t^A\), but \(z_t\) remains in the implementation formula for
\(f_t^*\). Allocation separation is therefore not institutional
independence between monetary and fiscal instruments. Appendix C.5 gives
the proof.

\subsection{Legacy wage dispersion and the post-impact RANK
benchmark}\label{sec-legacy-active}

The symmetric state is the special case in which active transfers
implement the full dual-rigidity RANK allocation.

\textbf{Corollary 1 (symmetric-state active-transfer implementation).}
If \(\sigma_{t-1}^w=0\), the policy authority can maintain \[
\sigma_t^w=\sigma_t^c=\sigma_t^n=\omega_t^A=0
\] and implement the dual-rigidity RANK allocation.

A nonzero legacy wage gap is different. From \(\sigma_{-1}^w\neq0\), the
policy authority can set \(\sigma_t^w=0\) for all \(t\ge0\), but doing
so cannot also set all impact-period distributional variables to zero.
The wage-dispersion equation requires an impact wedge, and the static
dispersion equation then creates an impact consumption gap.

Following Miyazaki (2026), I call this path the post-impact
RANK-implementation benchmark, or RI: \[
\sigma_t^w=0\quad \text{for all }t\ge0.
\] When \(\sigma_{-1}^w\neq0\), the impact values are \[
\omega_0^A=-\frac{1-\lambda}{\gamma\kappa_w}\sigma_{-1}^w,
\] and \[
\sigma_0^c=-\frac{1}{\gamma\kappa_w}\sigma_{-1}^w.
\] All distributional variables are zero from the next period onward. RI
is a useful implementation benchmark, but it is not the impact-period
RANK allocation and not a welfare optimum.

\textbf{Proposition 4 (Immediate post-impact RANK implementation is not
welfare optimal from a legacy state).} If \(\sigma_{-1}^w\neq0\), the
feasible RI path that sets \(\sigma_t^w=0\) for all \(t\ge0\) is not a
local optimum of the active-transfer distributional problem under the
maintained parameter restrictions.

Leaving a small part of the inherited wage gap in place reduces the
impact consumption gap and relative-wage adjustment cost at first order.
The resulting labor-dispersion and future adjustment costs increase only
at second order around RI. Appendix C.6 gives the feasible perturbation
and the loss derivative.

\textbf{Proposition 5 (Equivalence of active-transfer discretion and
commitment).} Under unrestricted targeted transfers, the CES-consistent
labor-dispersion weight, and the maintained restrictions \(0<\beta<1\),
\(\eta_w>0\), and \(W_n\equiv\varphi\psi_w^2+\psi_w>0\), Markov-perfect
discretion (AD) and date-0 commitment (AC) choose the same
distributional transition from any inherited wage gap. Let \[
\mathcal K=1+\beta+\frac{W_n}{\eta_w},
\] and let \[
F^*=\frac{2}{\mathcal K+\sqrt{\mathcal K^2-4\beta}}
\] be the stable root. Both policies choose \[
\sigma_t^w=(F^*)^{t+1}\sigma_{-1}^w,
\qquad
\sigma_t^c=0.
\] The unscaled distributional value coefficient is
\(P^*=\eta_w(1-F^*)\), so \(V^{dist}(d)=\lambda(1-\lambda)P^*d^2/2\) for
an inherited gap \(d\). Thus the infinite-horizon AD--AC loss difference
is exactly zero; finite-horizon numerical differences reflect terminal
truncation and solver roundoff. Appendix C.7 gives the proof.

\section{Quantitative Analysis}\label{sec-quantitative}

The numerical analysis first evaluates policy dependence and welfare
comparisons under passive transfers, then illustrates optimal adjustment
under active transfers. The passive-transfer exercises compare policy
regimes in the full model. The active-transfer experiment holds
aggregate allocation fixed and compares discounted distributional losses
along a transition from an inherited wage gap.

\subsection{Passive transfers}\label{sec-passive-quantitative}

Let \(\bar\ell\) denote the stationary unconditional expected period
loss, \(E_{\mathrm{stat}}[\ell_t]\), with a policy superscript
identifying the regime. Discounted transition losses from a specified
initial state are denoted by \(\mathcal L\).

The quantitative exercises measure policy dependence and
benchmark-relative loss differences separately. The first holds other
primitive states fixed and varies inherited wage dispersion. The second
asks how closely current aggregate variables span the nominal-rate
policy function. The third decomposes the stationary loss difference
between discretion and fixed rules. A small projection residual need not
imply a small welfare cost, and a loss difference between policy regimes
does not identify the value of a single state variable.

All fixed rules, including the price-inflation-only benchmark, commit to
future feedback coefficients. Their welfare comparisons therefore differ
from the descriptive projection of a given discretionary equilibrium.
The structural RANK and sticky-price-only regimes are diagnostic
contrasts rather than feasible policy alternatives within the maintained
economy.

\subsubsection{Benchmark regimes and illustrative
parameterization}\label{benchmark-regimes-and-illustrative-parameterization}

The benchmark design compares optimal discretion with three
one-parameter price-inflation-rule economies. The baseline simple-rule
diagnostic compares Regime 1 with Regime 2, holding fixed the
dual-rigidity TANK structure and changing the policy regime. Regimes 3
and 4 are diagnostic counterfactuals. They use the same coefficient
\(\phi_\pi^\star\) chosen from Regime 2, but remove either heterogeneity
or wage rigidity.

\begin{itemize}
\tightlist
\item
  Regime 1: optimal discretion (OD) in dual-rigidity TANK, the
  no-commitment allocation
\item
  Regime 2: optimized price-inflation-rule dual-rigidity TANK, the
  baseline simple-rule diagnostic
\item
  Regime 3: price-inflation-rule dual-rigidity RANK, a diagnostic
  benchmark without heterogeneity
\item
  Regime 4: price-inflation-rule sticky-price-only TANK, a diagnostic
  benchmark without wage rigidity
\end{itemize}

\Needspace{28\baselineskip}

\begin{longtable}[]{@{}
  >{\raggedright\arraybackslash}p{(\linewidth - 4\tabcolsep) * \real{0.2200}}
  >{\raggedleft\arraybackslash}p{(\linewidth - 4\tabcolsep) * \real{0.2000}}
  >{\raggedleft\arraybackslash}p{(\linewidth - 4\tabcolsep) * \real{0.5800}}@{}}
\caption{Illustrative benchmark
parameterization}\label{tbl-calibration}\tabularnewline
\toprule\noalign{}
\begin{minipage}[b]{\linewidth}\raggedright
Parameter
\end{minipage} & \begin{minipage}[b]{\linewidth}\raggedleft
Value
\end{minipage} & \begin{minipage}[b]{\linewidth}\raggedleft
Definition
\end{minipage} \\
\midrule\noalign{}
\endfirsthead
\toprule\noalign{}
\begin{minipage}[b]{\linewidth}\raggedright
Parameter
\end{minipage} & \begin{minipage}[b]{\linewidth}\raggedleft
Value
\end{minipage} & \begin{minipage}[b]{\linewidth}\raggedleft
Definition
\end{minipage} \\
\midrule\noalign{}
\endhead
\bottomrule\noalign{}
\endlastfoot
\(\beta\) & 0.99 & Discount factor \\
\(\gamma\) & 1.00 & Inverse intertemporal elasticity \\
\(\varphi\) & 1.00 & Inverse Frisch elasticity \\
\(\lambda\) & 0.25 & Hand-to-mouth share \\
\(\psi_p,\eta_p\) & 10.00, 50.00 & Goods substitution; price adjustment
cost \\
\(\psi_w,\eta_w\) & 10.00, 50.00 & Labor substitution; wage adjustment
cost \\
\(\phi_\pi^\star\) & 3.01 & Optimized price-inflation response \\
\(\rho_a,\rho_z\) & 0.80, 0.80 & Shock persistence \\
\(\sigma_a,\sigma_z\) & 0.01, 0.01 & Innovation standard deviations \\
\end{longtable}

\emph{Notes:} Both shocks have quarterly persistence 0.8 and innovation
standard deviation 0.01. The implied slopes are
\(\kappa_p=\kappa_w=0.2\). The price-only response coefficient is chosen
within \([1.01,5]\); it is not an externally calibrated policy
parameter.

The price-inflation rule is only a benchmark. The discretionary problem
in Section~\ref{sec-passive-policy} instead optimizes over feasible
allocations without a fixed feedback rule. The baseline parameterization
illustrates the preference, heterogeneity, and nominal-rigidity
structure of the dual-rigidity TANK canonical representation; it is not
an empirical fit to a particular economy.\footnote{ The benchmark values
  for nominal rigidities are chosen to preserve persistent but stable
  relative-wage adjustment and are in line with sticky-wage TANK
  calibrations such as Colciago (2011) and Ida and Okano (2024).}

Table~\ref{tbl-calibration} fixes the scale of the experiment. The
hand-to-mouth share is nontrivial, \(\lambda=0.25\), and price and wage
rigidity are symmetric, \(\eta_p=\eta_w=50\), so both aggregate nominal
adjustment and relative-wage adjustment can matter. With
\(\psi_p=\psi_w=10\), the implied Phillips-curve slopes are
\(\kappa_p=\kappa_w=0.2\). The coefficient in the price-inflation rule
is not calibrated externally; it minimizes the dual-rigidity TANK
benchmark loss over \(\phi_\pi\in[1.01,5]\). Online Appendix B.2 gives
the optimization details. Differences across Regime 1-Regime 4 come from
the policy rule or from shutting down a structural block, not from
changing the shock processes. The online appendix reports the solution
methods, supplementary results, and numerical diagnostics.

\subsubsection{State dependence and shock
responses}\label{state-dependence-and-shock-responses}

The direct response to the inherited wage gap holds the other primitive
states fixed. With the state ordered as
\(s_t=(w_{t-1},\sigma_{t-1}^w,a_t,z_t)^\top\), the baseline nominal-rate
coefficient is \[
\left.\frac{\partial i_t}{\partial\sigma_{t-1}^w}\right|_{w_{t-1},a_t,z_t}
=0.327577.
\] The corresponding real-rate coefficient is \(0.328516\). Thus a 1
percent inherited wage gap raises the nominal rate by about \(0.00328\)
in quarterly log-rate units (about 33 basis points in the net quarterly
rate to first order, without annualization) and lowers the output gap by
\(0.000107\), relative to the zero state. This is a local comparative
static of the solved discretionary policy.

\begin{figure}

\centering{

\includegraphics[width=1\linewidth,height=\textheight,keepaspectratio]{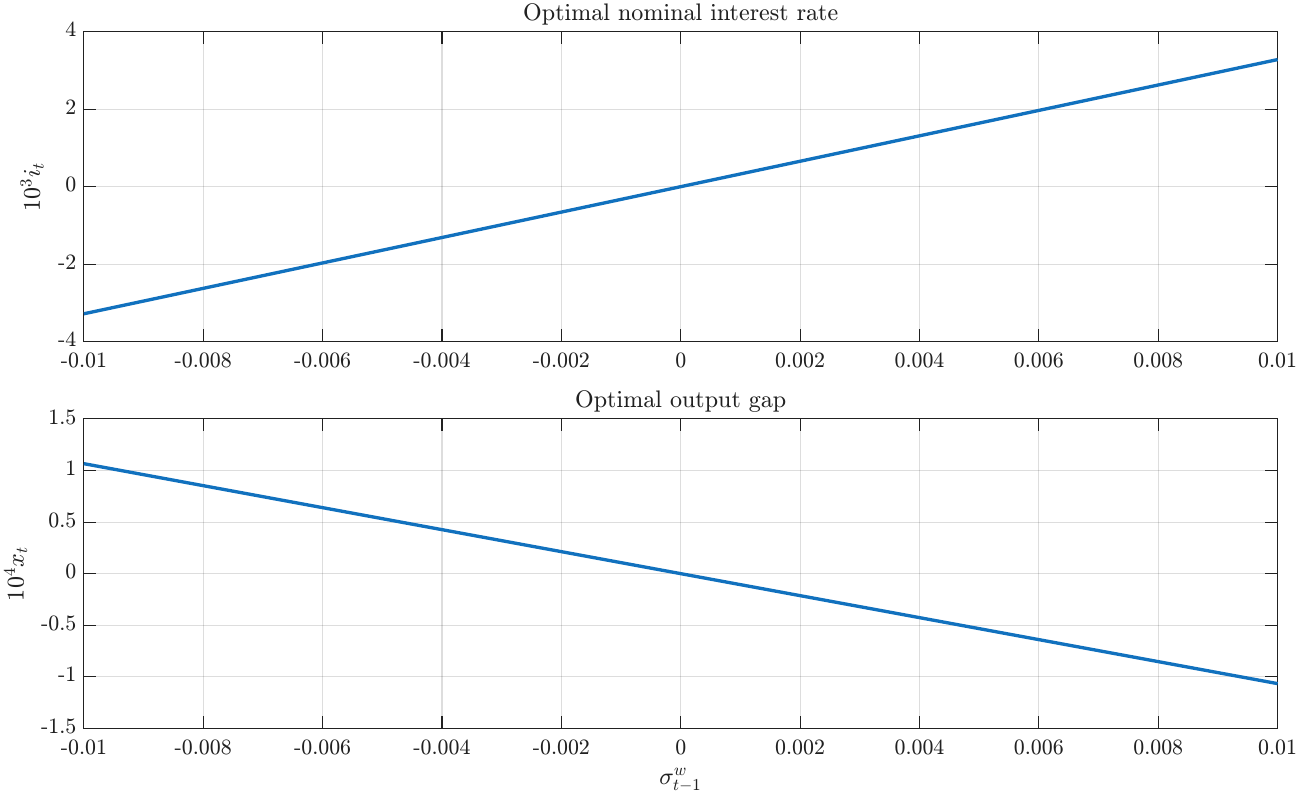}

\begin{minipage}{\linewidth}\raggedright

\emph{Notes:} The inherited wage gap varies over \([-0.01,0.01]\), with
the lagged aggregate real wage and both exogenous shocks held at zero.
The panels report the nominal interest rate and output gap. Rates are
quarterly log deviations, without annualization; all other variables are
log deviations.

\end{minipage}

}

\caption{\label{fig-policy-function}Discretionary responses to an
inherited wage gap}

\end{figure}%

A separate experiment starts from \(s_0=(0,0.01,0,0)'\) and sets all
future innovations to zero. Its discounted loss is
\(3.856\times10^{-4}\) under discretion, compared with
\(4.184\times10^{-4}\) under the baseline price-only rule. The positive
initial-period lower bound in Proposition 1 is \(3.223\times10^{-4}\).
These losses establish the cost of a pure legacy state; the difference
between the two policies still does not isolate the welfare value of
observing that state. Online Appendix B.5 reports the full paths and
loss accounting.

At the baseline, \(H\) has rank three and the stationary state
covariance is positive definite. A joint state perturbation with
\(\Delta\sigma_{t-1}^w=-0.01\) keeps \(H\Delta s=0\) to numerical
precision but changes the implementing nominal rate by approximately
\(I\Delta s=-0.00352549\). Thus the solved equilibrium satisfies the
conditions of Proposition 2. Online Appendix B.3 reports the full
perturbation and rank diagnostics.

The conditional projection answers a different question. Current price
inflation explains 66.45 percent of nominal-rate variance under the
baseline stationary covariance. Adding wage inflation raises this share
to 87.97 percent, and adding the output gap raises it to 99.61 percent.
Inherited wage dispersion then removes the remaining 0.3904 percentage
points of total variance. This is 100 percent of the residual variance,
but a small increment in total explanatory power. At the baseline, each
of the other primitive states can also complete the four-dimensional
span. The additional state is required for an exact aggregate-variable
policy representation under the rank conditions of Proposition 2; the
projection alone does not identify a unique quantitative contribution of
wage dispersion.

\begin{figure}

\centering{

\includegraphics[width=1\linewidth,height=\textheight,keepaspectratio]{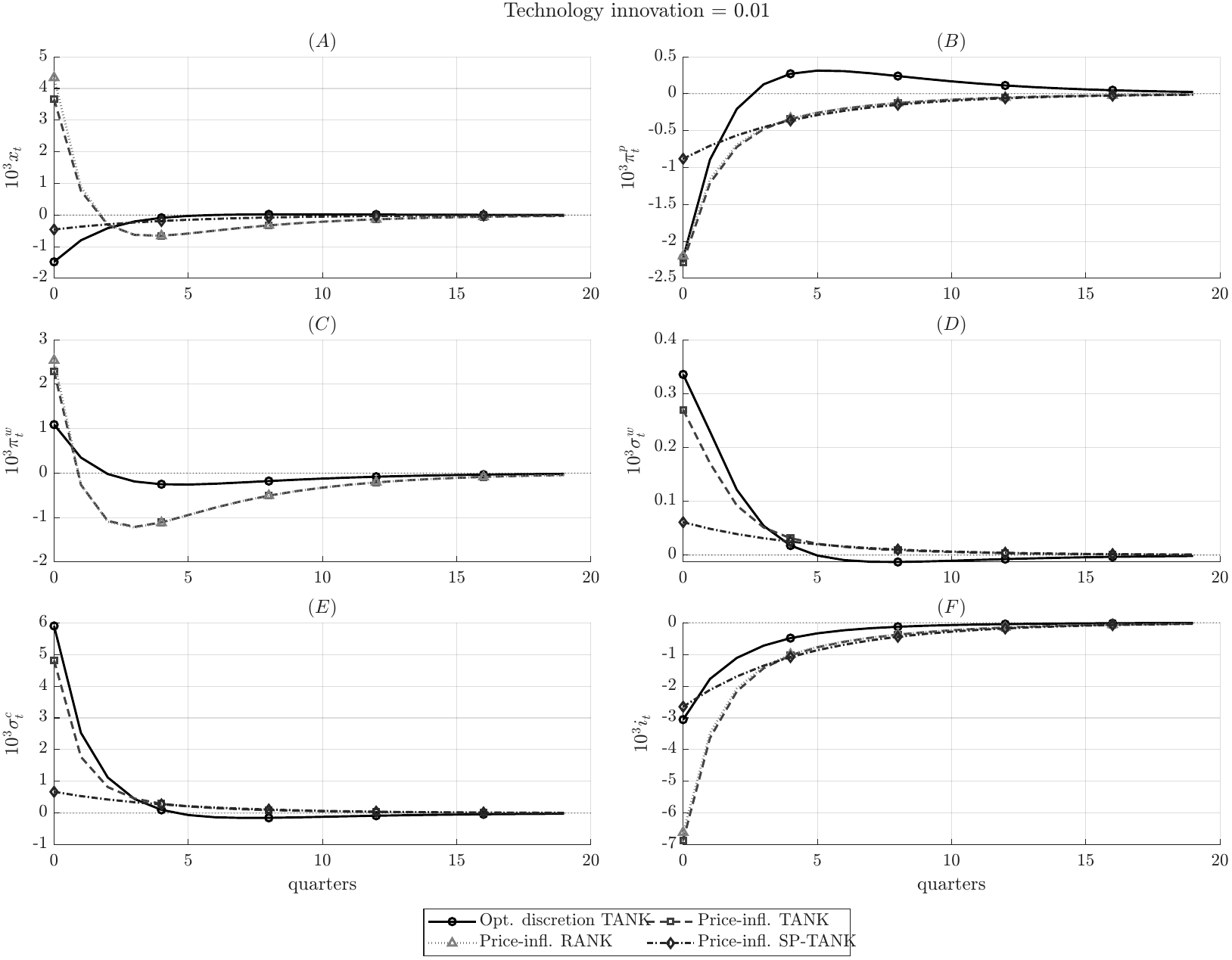}

\begin{minipage}{\linewidth}\raggedright

\emph{Notes:} A positive one-standard-deviation technology innovation is
0.01 in log units. Responses cover 20 quarters. R1 is discretion; R2 is
the optimized price-only rule in dual-rigidity TANK. R3 and R4 are
structural diagnostic regimes using the R2 coefficient. Interest rates
are nominal quarterly log deviations, without annualization; other
variables are log deviations.

\end{minipage}

}

\caption{\label{fig-technology-irf}Responses to a technology shock}

\end{figure}

\begin{figure}

\centering{

\includegraphics[width=1\linewidth,height=\textheight,keepaspectratio]{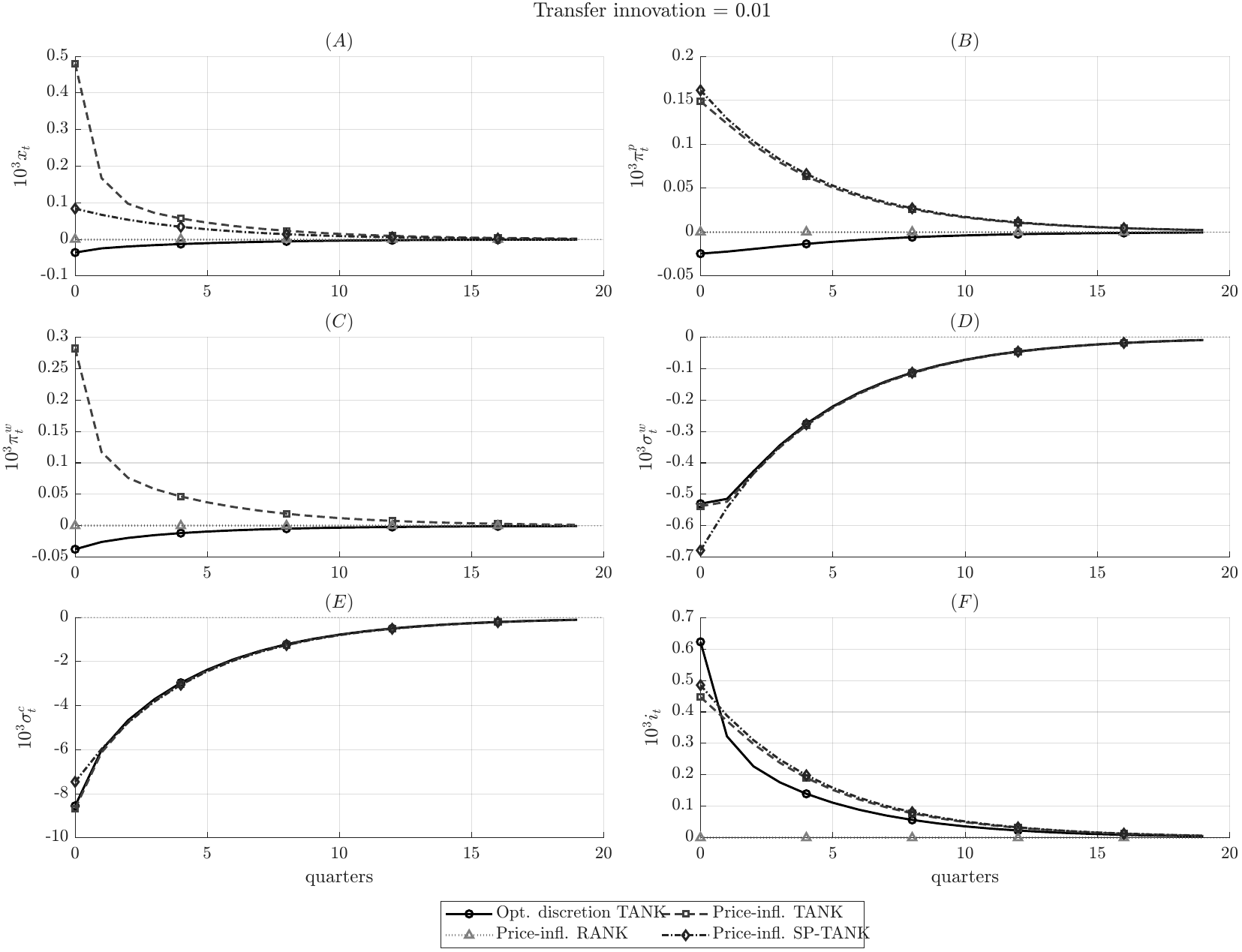}

\begin{minipage}{\linewidth}\raggedright

\emph{Notes:} A positive one-standard-deviation transfer innovation is
0.01, normalized by steady-state consumption. Responses cover 20
quarters, with regime definitions as in Figure~\ref{fig-technology-irf}.
The RANK response is zero because the shock has no cross-type transfer
margin in that economy. Interest rates are nominal quarterly log
deviations, without annualization; other variables are log deviations.

\end{minipage}

}

\caption{\label{fig-transfer-irf}Responses to a passive transfer shock}

\end{figure}

Technology and transfer shocks illustrate the different transmission
channels. Figure~\ref{fig-technology-irf} and
Figure~\ref{fig-transfer-irf} compare the same shocks across the four
regimes.

Relative to R2, discretion approximately halves the peak absolute
wage-inflation response to a technology shock, from \(2.3\times10^{-3}\)
to \(1.1\times10^{-3}\). Both TANK regimes also display substantial
consumption dispersion. These simultaneous aggregate and distributional
responses prevent the technology-shock comparison from isolating the
legacy wage-gap channel.

Under discretion, the transfer shock moves consumption dispersion by
about \(8.5\times10^{-3}\) on impact while the impact output gap and
both inflation rates remain below \(4\times10^{-5}\) in absolute value.
R2 produces larger aggregate responses. The nominal rate rises on impact
even though price inflation is slightly negative. This response is
consistent with the distributional transmission mechanism, but comparing
the two equilibria remains a policy-regime comparison.

\Needspace{32\baselineskip}

\subsubsection{Welfare comparisons and their
interpretation}\label{welfare-comparisons-and-their-interpretation}

\Needspace{24\baselineskip}

\begingroup
\small

\begin{longtable}[]{@{}
  >{\raggedright\arraybackslash}p{(\linewidth - 8\tabcolsep) * \real{0.3600}}
  >{\raggedleft\arraybackslash}p{(\linewidth - 8\tabcolsep) * \real{0.1600}}
  >{\raggedleft\arraybackslash}p{(\linewidth - 8\tabcolsep) * \real{0.1600}}
  >{\raggedleft\arraybackslash}p{(\linewidth - 8\tabcolsep) * \real{0.1600}}
  >{\raggedleft\arraybackslash}p{(\linewidth - 8\tabcolsep) * \real{0.1600}}@{}}
\caption{Stationary welfare-loss
decomposition}\label{tbl-welfare-decomposition}\tabularnewline
\toprule\noalign{}
\begin{minipage}[b]{\linewidth}\raggedright
Loss component
\end{minipage} & \begin{minipage}[b]{\linewidth}\raggedleft
R1
\end{minipage} & \begin{minipage}[b]{\linewidth}\raggedleft
R2
\end{minipage} & \begin{minipage}[b]{\linewidth}\raggedleft
R3
\end{minipage} & \begin{minipage}[b]{\linewidth}\raggedleft
R4
\end{minipage} \\
\midrule\noalign{}
\endfirsthead
\toprule\noalign{}
\begin{minipage}[b]{\linewidth}\raggedright
Loss component
\end{minipage} & \begin{minipage}[b]{\linewidth}\raggedleft
R1
\end{minipage} & \begin{minipage}[b]{\linewidth}\raggedleft
R2
\end{minipage} & \begin{minipage}[b]{\linewidth}\raggedleft
R3
\end{minipage} & \begin{minipage}[b]{\linewidth}\raggedleft
R4
\end{minipage} \\
\midrule\noalign{}
\endhead
\bottomrule\noalign{}
\endlastfoot
Output gap \(x_t\) & \(3.053\times10^{-6}\) & \(1.622\times10^{-5}\) &
\(2.160\times10^{-5}\) & \(6.018\times10^{-7}\) \\
Price inflation \(\pi_t^p\) & \(1.579\times10^{-4}\) &
\(1.944\times10^{-4}\) & \(1.784\times10^{-4}\) &
\(5.564\times10^{-5}\) \\
Wage inflation \(\pi_t^w\) & \(4.203\times10^{-5}\) &
\(2.970\times10^{-4}\) & \(3.264\times10^{-4}\) & 0 \\
Consumption dispersion \(\sigma_t^c\) & \(1.987\times10^{-5}\) &
\(1.897\times10^{-5}\) & 0 & \(1.460\times10^{-5}\) \\
Labor dispersion \(\sigma_t^n\) & \(1.280\times10^{-5}\) &
\(1.247\times10^{-5}\) & 0 & \(1.327\times10^{-5}\) \\
Relative wage inflation \(\Delta\sigma_t^w\) & \(2.111\times10^{-6}\) &
\(1.916\times10^{-6}\) & 0 & 0 \\
Total & \(2.378\times10^{-4}\) & \(5.410\times10^{-4}\) &
\(5.264\times10^{-4}\) & \(8.411\times10^{-5}\) \\
\end{longtable}

\emph{Notes:} Entries are contributions to stationary expected period
loss \(\bar\ell\). R1 is optimal discretion; R2 is the optimized
price-inflation-only rule in the same dual-rigidity TANK economy. R3 is
dual-rigidity RANK and R4 is sticky-price-only TANK, both using the R2
coefficient. R3 and R4 are structural diagnostic contrasts. They are not
feasible policy alternatives for the maintained economy.

\endgroup

The discretionary stationary loss is \(2.378\times10^{-4}\), compared
with \(5.410\times10^{-4}\) under the optimized price-only rule, a
56.052 percent reduction relative to that rule. Aggregate wage-inflation
stabilization accounts for 84.10 percent of the loss difference in
Table~\ref{tbl-welfare-decomposition}. In contrast, each of the three
cross-sectional loss components is slightly higher under discretion. The
comparison therefore provides little support for interpreting the total
reduction as the size of the inherited distributional-state channel.

\Needspace{16\baselineskip}

\begin{longtable}[]{@{}
  >{\raggedright\arraybackslash}p{(\linewidth - 6\tabcolsep) * \real{0.2500}}
  >{\raggedleft\arraybackslash}p{(\linewidth - 6\tabcolsep) * \real{0.2500}}
  >{\raggedleft\arraybackslash}p{(\linewidth - 6\tabcolsep) * \real{0.2500}}
  >{\raggedleft\arraybackslash}p{(\linewidth - 6\tabcolsep) * \real{0.2500}}@{}}
\caption{Loss decomposition by
shock}\label{tbl-shock-welfare}\tabularnewline
\toprule\noalign{}
\begin{minipage}[b]{\linewidth}\raggedright
Innovations
\end{minipage} & \begin{minipage}[b]{\linewidth}\raggedleft
OD loss
\end{minipage} & \begin{minipage}[b]{\linewidth}\raggedleft
Price-rule loss
\end{minipage} & \begin{minipage}[b]{\linewidth}\raggedleft
Reduction (\%)
\end{minipage} \\
\midrule\noalign{}
\endfirsthead
\toprule\noalign{}
\begin{minipage}[b]{\linewidth}\raggedright
Innovations
\end{minipage} & \begin{minipage}[b]{\linewidth}\raggedleft
OD loss
\end{minipage} & \begin{minipage}[b]{\linewidth}\raggedleft
Price-rule loss
\end{minipage} & \begin{minipage}[b]{\linewidth}\raggedleft
Reduction (\%)
\end{minipage} \\
\midrule\noalign{}
\endhead
\bottomrule\noalign{}
\endlastfoot
Technology & \(2.094\times10^{-4}\) & \(5.072\times10^{-4}\) & 58.71 \\
Transfer & \(2.834\times10^{-5}\) & \(3.379\times10^{-5}\) & 16.11 \\
Both & \(2.378\times10^{-4}\) & \(5.410\times10^{-4}\) & 56.05 \\
\end{longtable}

\emph{Notes:} OD denotes optimal discretion; losses are stationary
expected period losses \(\bar\ell\). Each row changes only the
innovation covariance. Both policy functions, including the price-rule
coefficient optimized under the baseline two-shock calibration, are held
fixed. The shocks are independent, so the two single-shock losses add to
the joint loss. The percentage reduction uses the price-rule loss in
that row as its denominator. This is a contribution decomposition, not a
shock-specific rule reoptimization.

Table~\ref{tbl-shock-welfare} holds both baseline policy functions fixed
and decomposes loss by independent innovations. Technology shocks
account for 98.21 percent of the total loss difference. The reduction is
58.71 percent for technology shocks alone and 16.11 percent for transfer
shocks alone. Even the latter compares complete policy regimes; it does
not measure the marginal welfare cost of removing wage-gap information.

\Needspace{16\baselineskip}

\begingroup
\small

\begin{longtable}[]{@{}
  >{\raggedright\arraybackslash}p{(\linewidth - 10\tabcolsep) * \real{0.2800}}
  >{\raggedleft\arraybackslash}p{(\linewidth - 10\tabcolsep) * \real{0.1100}}
  >{\raggedleft\arraybackslash}p{(\linewidth - 10\tabcolsep) * \real{0.1100}}
  >{\raggedleft\arraybackslash}p{(\linewidth - 10\tabcolsep) * \real{0.1100}}
  >{\raggedleft\arraybackslash}p{(\linewidth - 10\tabcolsep) * \real{0.2000}}
  >{\raggedleft\arraybackslash}p{(\linewidth - 10\tabcolsep) * \real{0.1900}}@{}}
\caption{Fixed rules with additional aggregate
responses}\label{tbl-rich-simple-rules}\tabularnewline
\toprule\noalign{}
\begin{minipage}[b]{\linewidth}\raggedright
Rule
\end{minipage} & \begin{minipage}[b]{\linewidth}\raggedleft
\(\phi_\pi\)
\end{minipage} & \begin{minipage}[b]{\linewidth}\raggedleft
\(\phi_w\)
\end{minipage} & \begin{minipage}[b]{\linewidth}\raggedleft
\(\phi_x\)
\end{minipage} & \begin{minipage}[b]{\linewidth}\raggedleft
Rule loss
\end{minipage} & \begin{minipage}[b]{\linewidth}\raggedleft
OD reduction (\%)
\end{minipage} \\
\midrule\noalign{}
\endfirsthead
\toprule\noalign{}
\begin{minipage}[b]{\linewidth}\raggedright
Rule
\end{minipage} & \begin{minipage}[b]{\linewidth}\raggedleft
\(\phi_\pi\)
\end{minipage} & \begin{minipage}[b]{\linewidth}\raggedleft
\(\phi_w\)
\end{minipage} & \begin{minipage}[b]{\linewidth}\raggedleft
\(\phi_x\)
\end{minipage} & \begin{minipage}[b]{\linewidth}\raggedleft
Rule loss
\end{minipage} & \begin{minipage}[b]{\linewidth}\raggedleft
OD reduction (\%)
\end{minipage} \\
\midrule\noalign{}
\endhead
\bottomrule\noalign{}
\endlastfoot
Price & 3.006 & 0.000 & 0.000 & \(5.410\times10^{-4}\) & 56.052 \\
Price and wage & 3.872 & 5.000 & 0.000 & \(2.413\times10^{-4}\) &
1.448 \\
Price, wage and output & 1.010 & 5.000 & 3.993 & \(2.330\times10^{-4}\)
& -2.030 \\
\end{longtable}

\emph{Notes:} OD denotes optimal discretion; losses are stationary
expected period losses \(\bar\ell\). The search box is
\(\phi_\pi\in[1.01,5]\) and \(\phi_w,\phi_x\in[0,5]\). The last column
is \(100(\bar\ell^{rule}-\bar\ell^{OD})/\bar\ell^{rule}\), so a negative
entry favors the fixed rule. Both richer specifications reach the
wage-response upper bound; the three-variable specification also reaches
the price-response lower bound. Every fixed rule assumes commitment to
future coefficients.

\endgroup

The richer fixed rules in Table~\ref{tbl-rich-simple-rules} show how
much of the price-only benchmark gap is associated with omitted
aggregate responses. Within the initial search box, adding wage
inflation reduces the discretionary advantage to 1.448 percent. Adding
the output gap produces a rule with 1.989 percent lower loss than
discretion when discretionary loss is the denominator. The table instead
uses rule loss as its denominator, giving \(-2.030\) percent in the
final column. All three rule classes commit to their future
coefficients.

Both richer searches reach the wage-response upper bound of 5, and the
three-variable rule also reaches the price-response lower bound.
Expanding the wage bound to 10 and then 20 slightly improves the best
losses found. The two-inflation rule then reaches the price-response
upper bound, while the three-variable rule reaches the output-response
upper bound. Online Appendix B.4 reports every bound and the resulting
coefficients. These are bounded numerical searches, not certified global
optima or substitutes for a full Ramsey comparison.

\subsubsection{Sensitivity and the scope of the
evidence}\label{sensitivity-and-the-scope-of-the-evidence}

\begin{figure}

\centering{

\includegraphics[width=1\linewidth,height=\textheight,keepaspectratio]{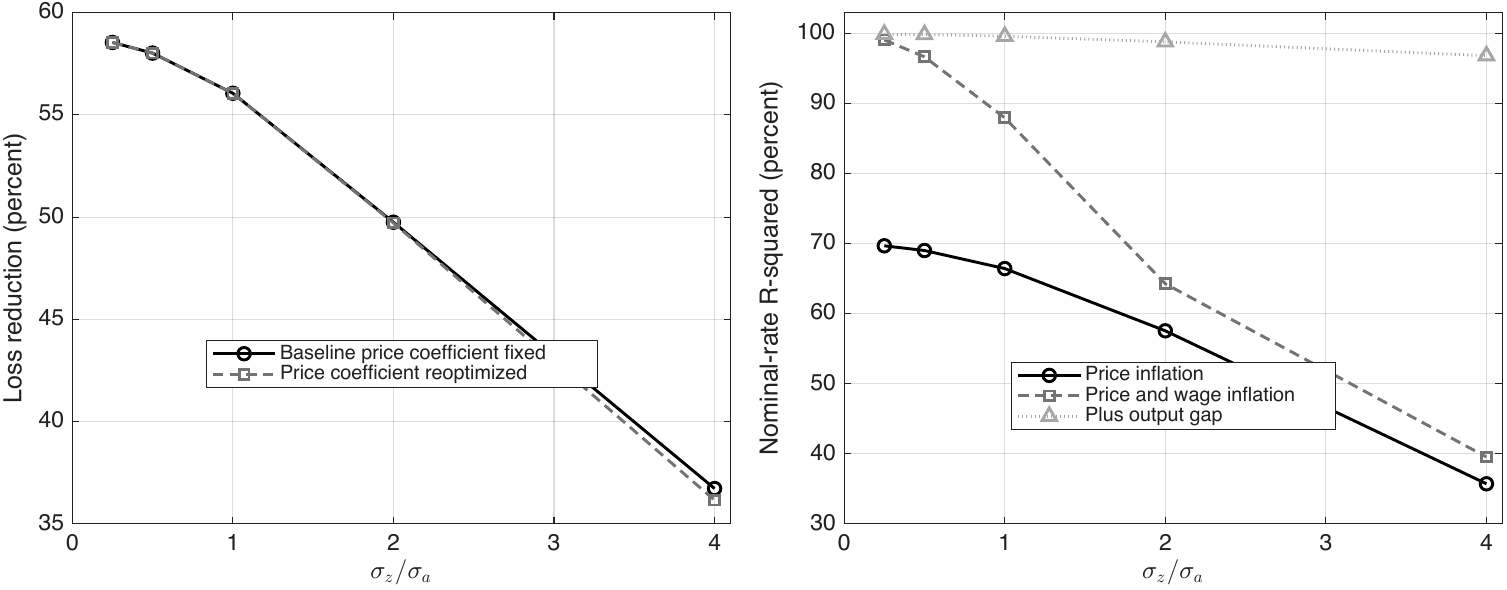}

\begin{minipage}{\linewidth}\raggedright

\emph{Notes:} The horizontal axis is the innovation standard-deviation
ratio \(\sigma_z/\sigma_a\), with \(\sigma_a=0.01\) and
\(\rho_a=\rho_z=0.8\). The left panel compares a fixed baseline price
coefficient with reoptimization at each ratio; each curve uses its own
rule loss as the denominator. The right panel projects the discretionary
nominal rate on price inflation alone, both inflation rates, and both
inflation rates plus the output gap.

\end{minipage}

}

\caption{\label{fig-shock-ratio}Sensitivity to the relative size of
transfer shocks}

\end{figure}

The relative shock scale affects both welfare comparisons and projection
fit. Figure~\ref{fig-shock-ratio} varies the transfer-to-technology
innovation standard-deviation ratio while holding persistence fixed. At
every point, it reoptimizes the price-only coefficient; the
discretionary policy coefficients are unchanged by the innovation
covariance in this linear-quadratic problem. Increasing the ratio from
0.25 to 4 reduces the price-rule-relative loss reduction from 58.53
percent to 36.16 percent. The projection also changes because it weights
state directions using the new stationary covariance.

Labor substitutability changes both the wage Phillips-curve slope,
\(\kappa_w=\psi_w/\eta_w\), and the labor-dispersion loss weight,
\(\varphi\psi_w^2+\psi_w\). Across \(\psi_w\in\{4,6,10,15,20\}\), the
direct nominal-rate response to inherited dispersion remains positive
but is not monotone. Neither is the three-aggregate projection fit: its
\(R^2\) ranges from 0.9616 to nearly one. Online Appendix B.6 reports
these results alongside the heterogeneity and adjustment-cost
sensitivities. A common sign of the direct response across this grid
does not imply a common quantitative importance under the stationary
shock distribution.

The exercises distinguish structural state relevance, linear spanning,
and policy-regime welfare differences. They do not solve a discretionary
equilibrium in which the policymaker cannot observe the wage gap. Such
an information restriction requires a specified observation process and
beliefs; setting one coefficient of the full-information solution to
zero would not define that equilibrium. The marginal welfare value of
distributional information remains outside the present comparison.

\subsection{Active transfers}\label{sec-active-commitment}

The active-transfer calculation illustrates the theoretical results in
Section~\ref{sec-active-transfers} using the same structural parameter
values as Table~\ref{tbl-calibration}. It is a reduced
distribution-block diagnostic. It fixes all aggregate variables at zero
and sets the only initial state to \(\sigma_{-1}^w=0.01\). It compares
AD, AC, and RI as defined in Section~\ref{sec-legacy-active}. AC is not
a full Ramsey policy for the whole economy, and the calculation excludes
the usual commitment gain in the aggregate dual-rigidity RANK block.

\begin{figure}

\centering{

\includegraphics[width=1\linewidth,height=\textheight,keepaspectratio]{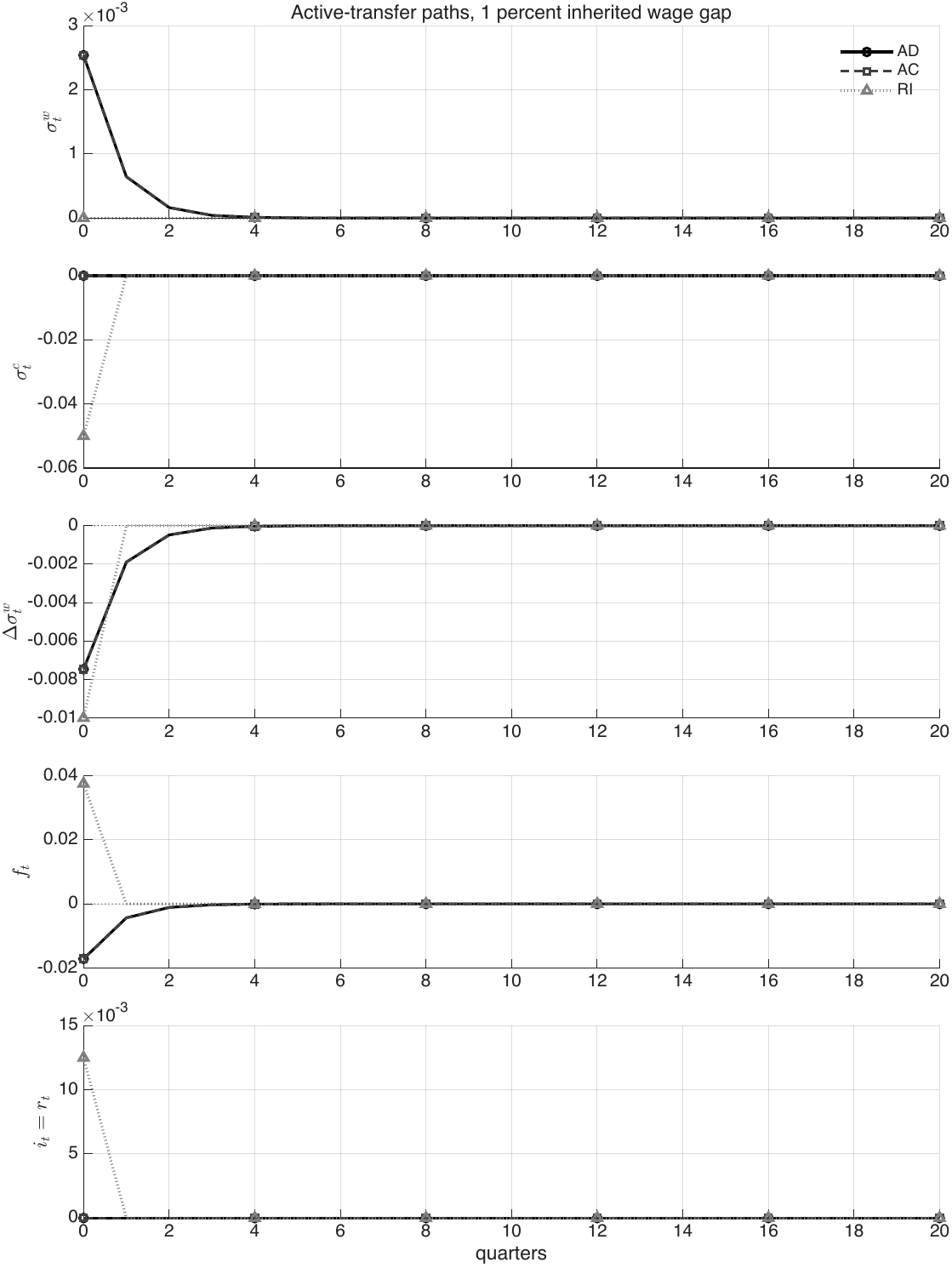}

\begin{minipage}{\linewidth}\raggedright

\emph{Notes:} The initial wage gap is 0.01 and aggregate variables and
shocks are zero. AD denotes discretion, AC date-0 commitment in the
distribution block, and RI immediate post-impact wage-gap elimination.
AD and AC coincide. Rates are nominal quarterly log deviations, without
annualization; they coincide with real rates because expected price
inflation is zero. The rate implements the fixed aggregate allocation.
Other variables are log deviations, with transfers normalized by
steady-state consumption.

\end{minipage}

}

\caption{\label{fig-active-transfer-paths}Active-transfer adjustment
from an inherited wage gap}

\end{figure}%

Figure~\ref{fig-active-transfer-paths} shows the exact AD-AC equivalence
implied by the CES-consistent objective. Both leave about 25.4 percent
of the inherited wage gap in place on impact and then converge over the
next few quarters. RI eliminates the wage gap immediately, but it
requires a much larger impact consumption gap and larger implementing
policy movements.

The common discounted distributional loss under AD and AC, divided by
the squared initial gap \((\sigma_{-1}^w)^2\), is 3.49737, compared with
7.03125 under RI. Optimal smoothing therefore reduces loss by 50.260
percent relative to RI, using RI loss as the denominator. This
comparison does not measure the gain from replacing passive transfers
with active transfers. Consumption dispersion is zero on the optimal
transition; labor dispersion accounts for 20.30 percent of loss and
relative-wage adjustment for 79.70 percent. Online Appendix B.9 reports
the full decomposition and numerical diagnostics.

The separation result in Section~\ref{sec-active-transfers} is
analytical; it does not depend on the calculation. The RI-AD gap
measures the dynamic distributional cost of forcing an immediate
post-impact RANK implementation from a nonzero legacy state. The exact
AD-AC equivalence is specific to the CES-consistent separated
distribution block. It does not characterize full-model Ramsey gains,
cost-push shocks, aggregate price and wage stabilization under
commitment, public debt, distortionary taxation, transfer limits,
transfer adjustment costs, or the zero lower bound.

\section{Conclusion}\label{sec-conclusion}

Type-specific own-lag wage adjustment adds a distributional state to the
aggregate real-wage state already present under dual rigidity. With
passive transfers, inherited wage dispersion changes implementable
allocations and has a positive welfare cost. Its effect on the nominal
interest rate is a separate question: explicit rank conditions
characterize when current price inflation, wage inflation, and the
output gap fail to determine the implementing rate. The baseline
equilibrium satisfies those conditions.

The quantitative evidence also limits the interpretation of that result.
The three aggregate variables explain almost all stationary nominal-rate
variation. The large loss reduction relative to an optimized
price-inflation-only rule is driven mainly by aggregate wage-inflation
stabilization following technology shocks, and richer fixed rules
substantially close or reverse that comparison. These exercises
establish policy dependence and characterize specified benchmarks; they
do not estimate the marginal welfare cost of withholding wage-gap
information from a discretionary policymaker.

Unrestricted targeted transfers change the allocation problem. Aggregate
stabilization separates from the legacy wage-gap transition, while the
cost of relative-wage adjustment remains. Immediate representative-agent
implementation is inefficient from a nonzero inherited gap. Under the
maintained CES-consistent distributional objective, optimal smoothing
eliminates consumption dispersion and is time consistent: discretion and
date-0 commitment select the same transition. The institutional
distinction is therefore which policy instrument manages the inherited
state, and which adjustment costs remain after aggregate separation.

\section*{References}\label{references}
\addcontentsline{toc}{section}{References}

\protect\phantomsection\label{refs}

\section*{Appendices}\label{appendices}
\addcontentsline{toc}{section}{Appendices}

\hyperref[sec-appendix-a]{Appendix A} derives the second-order objective
implied by the maintained wage-setting structure.
\hyperref[sec-appendix-b]{Appendix B} gives the reduced recursive
discretionary system used in the main text.
\hyperref[sec-appendix-c]{Appendix C} contains the proposition proofs
and theoretical benchmark derivations, including the active-transfer
separation result. The online appendix gives the full Lagrangian and
first-order-condition derivation in Part A and the numerical methods and
supplementary evidence in Part B, including policy projections,
richer-rule searches, sensitivity analysis, and the active-transfer
calculation.

\section*{Appendix A. Derivation of the Period Objective
Function}\label{sec-appendix-a}
\addcontentsline{toc}{section}{Appendix A. Derivation of the Period
Objective Function}

This appendix derives the quadratic period objective used in the main
text. I expand household utility to second order, rewrite the quadratic
terms in aggregate and dispersion form, use the resource constraint to
introduce nominal-adjustment costs, and then express the aggregate
component in output-gap form.

Two distinctions matter. First, the correct second-order aggregation
formulas must respect the CES labor aggregator and involve dispersion
terms rather than raw quadratic means. Second, the welfare-based
wage-adjustment term depends on cross-type wage-inflation dispersion.
Under the maintained own-lag wage-inflation specification, this object
is the change in the cross-type wage gap.

The derivation leads to the period loss \[
\begin{aligned}
\ell_t
={}&
\frac{1}{2}
\left\{
(\gamma+\varphi)x_t^2 + \eta_p(\pi_t^p)^2 + \eta_w(\pi_t^w)^2
\right\} \\
&+
\frac{\lambda(1-\lambda)}{2}
\left\{
\gamma(\sigma_t^c)^2 + \left(\varphi+\frac{1}{\psi_w}\right)(\sigma_t^n)^2 + \eta_w(\sigma_t^w-\sigma_{t-1}^w)^2
\right\}.
\end{aligned}
\] The steps below show where each term comes from and which terms are
independent of policy.

Throughout, \(\text{t.i.p.}\) denotes terms independent of policy, and
\(O(3)\) denotes terms of order three or higher.

\subsection*{A.1 Household-level welfare
expansion}\label{a.1-household-level-welfare-expansion}
\addcontentsline{toc}{subsection}{A.1 Household-level welfare expansion}

Let the period utility of household type \(j \in \{H,S\}\) be \[
U_t^j = \frac{(C_t^j)^{1-\gamma}}{1-\gamma} - \frac{(N_t^j)^{1+\varphi}}{1+\varphi}.
\] Define the period social welfare criterion as the population-weighted
sum of deviations from steady-state utility, \[
\mathbb{W}_t = \lambda (U_t^H-U^H) + (1-\lambda)(U_t^S-U^S).
\] Normalize by the steady-state marginal-utility scale \(U_C C\), where
\(U_C=C^{-\gamma}\) is the marginal utility of consumption at the
symmetric steady state. The efficient symmetric steady state derived in
Section~\ref{sec-model-foundations} gives \(C=N=1\) and
\(N^{1+\varphi}/C^{1-\gamma}=1\). This efficiency condition justifies
the unit coefficient on labor below; symmetry gives the same
normalization across types. Using log deviations \[
c_t^j = \log\left(\frac{C_t^j}{C}\right),
\qquad
n_t^j = \log\left(\frac{N_t^j}{N}\right),
\] a second-order Taylor expansion gives \[
\frac{U_t^j-U^j}{U_C C}
=
c_t^j + \frac{1-\gamma}{2}(c_t^j)^2
-
n_t^j - \frac{1+\varphi}{2}(n_t^j)^2
+ O(3).
\] Hence normalized welfare can be written as \[
\widetilde{\mathbb{W}}_t
\equiv
\frac{\mathbb{W}_t}{U_C C}
=
\bar c_t - \bar n_t
+ \frac{1-\gamma}{2}Q_t^c
- \frac{1+\varphi}{2}Q_t^n
+ O(3),
\] where \[
\bar c_t \equiv \lambda c_t^H + (1-\lambda)c_t^S,
\qquad
Q_t^c \equiv \lambda (c_t^H)^2 + (1-\lambda)(c_t^S)^2,
\] \[
\bar n_t \equiv \lambda n_t^H + (1-\lambda)n_t^S,
\qquad
Q_t^n \equiv \lambda (n_t^H)^2 + (1-\lambda)(n_t^S)^2.
\]

\subsection*{A.2 Aggregation and cross-sectional
decomposition}\label{a.2-aggregation-and-cross-sectional-decomposition}
\addcontentsline{toc}{subsection}{A.2 Aggregation and cross-sectional
decomposition}

Define aggregate log quantities and cross-type dispersion as \[
c_t = \log\left(\frac{C_t}{C}\right),
\qquad
n_t = \log\left(\frac{N_t}{N}\right),
\] \[
\sigma_t^c = c_t^S-c_t^H,
\qquad
\sigma_t^n = n_t^S-n_t^H.
\] Consumption is aggregated arithmetically, while labor services are
aggregated with the CES technology used in the model. Thus the exact
level relations are \[
\frac{C_t}{C} = \lambda e^{c_t^H} + (1-\lambda)e^{c_t^S},
\qquad
\frac{N_t}{N}
=
\left[
\lambda e^{\rho_w n_t^H}+(1-\lambda)e^{\rho_w n_t^S}
\right]^{1/\rho_w},
\qquad
\rho_w\equiv 1-\frac{1}{\psi_w},
\] with the Cobb--Douglas case \(\rho_w=0\) understood by continuity.
The corresponding second-order log approximations are \[
c_t = \bar c_t + \frac{1}{2}\lambda(1-\lambda)(\sigma_t^c)^2 + O(3),
\] \[
n_t = \bar n_t + \frac{\rho_w}{2}\lambda(1-\lambda)(\sigma_t^n)^2 + O(3).
\] Likewise, \[
Q_t^c = c_t^2 + \lambda(1-\lambda)(\sigma_t^c)^2 + O(3),
\] \[
Q_t^n = n_t^2 + \lambda(1-\lambda)(\sigma_t^n)^2 + O(3).
\] Substituting these relations into the household-level welfare
expansion yields \[
\widetilde{\mathbb{W}}_t
=
c_t - n_t
+ \frac{1-\gamma}{2}c_t^2
- \frac{1+\varphi}{2}n_t^2
- \frac{\lambda(1-\lambda)}{2}
\left\{
\gamma(\sigma_t^c)^2 + \left(\varphi+\frac{1}{\psi_w}\right)(\sigma_t^n)^2
\right\}
+ O(3).
\] This step gives the cross-type consumption-dispersion and
labor-dispersion penalties that appear in the policy objective. The
additional \(1/\psi_w\) in the labor-dispersion coefficient is the
curvature contribution from CES labor aggregation.

\subsection*{A.3 Resource constraint, production, and nominal-adjustment
costs}\label{a.3-resource-constraint-production-and-nominal-adjustment-costs}
\addcontentsline{toc}{subsection}{A.3 Resource constraint, production,
and nominal-adjustment costs}

The production technology is linear, so in logs the labor requirement is
exact: \[
n_t = y_t-a_t.
\] Let \(\pi_t^p\) denote aggregate price inflation and let \(\pi_t^H\)
and \(\pi_t^S\) denote type-specific wage inflation. Around the
zero-inflation steady state, the aggregate resource constraint implies
\[
c_t
=
y_t
- \frac{\eta_p}{2}(\pi_t^p)^2
- \frac{\eta_w}{2}
\left\{
\lambda(\pi_t^H)^2 + (1-\lambda)(\pi_t^S)^2
\right\}
+ O(3).
\] Because the adjustment-cost terms are already second order, they can
be ignored inside the quadratic terms \(c_t^2\) and \(n_t^2\). Therefore
\[
c_t - n_t
=
a_t
- \frac{\eta_p}{2}(\pi_t^p)^2
- \frac{\eta_w}{2}
\left\{
\lambda(\pi_t^H)^2 + (1-\lambda)(\pi_t^S)^2
\right\}
+ O(3),
\] \[
c_t^2 = y_t^2 + O(3),
\qquad
n_t^2 = (y_t-a_t)^2 + O(3).
\] Substituting these expressions into normalized welfare gives \[
\begin{aligned}
\widetilde{\mathbb{W}}_t
={}&
a_t
+ \frac{1-\gamma}{2}y_t^2
- \frac{1+\varphi}{2}(y_t-a_t)^2 \\
&- \frac{\eta_p}{2}(\pi_t^p)^2
- \frac{\eta_w}{2}
\left\{
\lambda(\pi_t^H)^2 + (1-\lambda)(\pi_t^S)^2
\right\} \\
&- \frac{\lambda(1-\lambda)}{2}
\left\{
\gamma(\sigma_t^c)^2 + \left(\varphi+\frac{1}{\psi_w}\right)(\sigma_t^n)^2
\right\}
+ O(3).
\end{aligned}
\] Expanding the aggregate part yields \[
\begin{aligned}
a_t
+ \frac{1-\gamma}{2}y_t^2
- \frac{1+\varphi}{2}(y_t-a_t)^2
={}&
a_t
+ (1+\varphi)a_t y_t \\
&- \frac{\gamma+\varphi}{2}y_t^2
- \frac{1+\varphi}{2}a_t^2.
\end{aligned}
\] The wage-adjustment term separates into an aggregate part and a
cross-sectional part. Evaluating inflation to first order, define
aggregate wage inflation by \[
\pi_t^w \equiv \lambda \pi_t^H + (1-\lambda)\pi_t^S.
\] For these first-order inflation terms, the exact quadratic
decomposition is \[
\lambda(\pi_t^H)^2 + (1-\lambda)(\pi_t^S)^2
=
(\pi_t^w)^2 + \lambda(1-\lambda)(\pi_t^S-\pi_t^H)^2.
\] Under the own-lag wage-inflation identity used in the paper, \[
\pi_t^j = w_t^j - w_{t-1}^j + \pi_t^p,
\qquad j \in \{H,S\},
\] so the cross-type wage-inflation dispersion is related to the
wage-gap state by \[
\pi_t^S-\pi_t^H = \sigma_t^w - \sigma_{t-1}^w.
\] The second-order welfare-based period objective is \[
\begin{aligned}
\widetilde{\mathbb{W}}_t
={}&
a_t + (1+\varphi)a_t y_t
- \frac{\gamma+\varphi}{2}y_t^2
- \frac{1+\varphi}{2}a_t^2 \\
&- \frac{\eta_p}{2}(\pi_t^p)^2
- \frac{\eta_w}{2}(\pi_t^w)^2
- \frac{\lambda(1-\lambda)}{2}\eta_w(\sigma_t^w-\sigma_{t-1}^w)^2 \\
&- \frac{\lambda(1-\lambda)}{2}
\left\{
\gamma(\sigma_t^c)^2 + \left(\varphi+\frac{1}{\psi_w}\right)(\sigma_t^n)^2
\right\}
+ O(3).
\end{aligned}
\]

\subsection*{A.4 Output-gap
representation}\label{a.4-output-gap-representation}
\addcontentsline{toc}{subsection}{A.4 Output-gap representation}

Define natural output and the output gap as \[
y_t^f = \frac{1+\varphi}{\gamma+\varphi}a_t,
\qquad
x_t = y_t-y_t^f.
\] Completing the square gives \[
a_t + (1+\varphi)a_t y_t
- \frac{\gamma+\varphi}{2}y_t^2
- \frac{1+\varphi}{2}a_t^2
=
-\frac{\gamma+\varphi}{2}x_t^2 + \text{t.i.p.}
\] Welfare can then be written as \[
\widetilde{\mathbb{W}}_t
=
-\ell_t + \text{t.i.p.} + O(3),
\] where the welfare-based quadratic period loss is \[
\begin{aligned}
\ell_t
={}&
\frac{1}{2}
\left\{
(\gamma+\varphi)x_t^2 + \eta_p(\pi_t^p)^2 + \eta_w(\pi_t^w)^2
\right\} \\
&+
\frac{\lambda(1-\lambda)}{2}
\left\{
\gamma(\sigma_t^c)^2 + \left(\varphi+\frac{1}{\psi_w}\right)(\sigma_t^n)^2 + \eta_w(\sigma_t^w-\sigma_{t-1}^w)^2
\right\}.
\end{aligned}
\] This quadratic loss is the maintained second-order approximation
around the efficient symmetric steady state, under the stated
wage-setting and fiscal closure. The corresponding discounted policy
objective is \[
\mathcal L
=
E_0\sum_{t=0}^{\infty}\beta^t\ell_t.
\]

\subsection*{A.5 State-control interpretation used in the recursive
problem}\label{a.5-state-control-interpretation-used-in-the-recursive-problem}
\addcontentsline{toc}{subsection}{A.5 State-control interpretation used
in the recursive problem}

The quadratic loss derived above is already in the state-control form
needed for the recursive discretionary problem. The inherited state
\(\sigma_{t-1}^w\) enters directly through the term
\(\sigma_t^w-\sigma_{t-1}^w\), which is cross-type wage-inflation
dispersion under the own-lag wage-inflation identity. Thus inherited
wage dispersion affects discretion both through the law of motion for
\(\sigma_t^w\) and through the current-period welfare criterion.

\section*{Appendix B. Reduced Recursive Discretionary
System}\label{sec-appendix-b}
\addcontentsline{toc}{section}{Appendix B. Reduced Recursive
Discretionary System}

This appendix reports the recursive system used in the main text. The
online appendix gives the full Lagrangian, first-order conditions,
static eliminations, envelope conditions, and implementation notes.

The wage-adjustment component of the second-order period loss penalizes
cross-type wage-inflation dispersion. The labor-dispersion component
separately penalizes the wage-gap level. Under the own-lag
wage-inflation identity, this term equals the change in the wage-gap
state. The inherited wage-gap state therefore enters through the
constraints and directly through the objective.

\subsection*{B.1 Recursive setup and Bellman
problem}\label{b.1-recursive-setup-and-bellman-problem}
\addcontentsline{toc}{subsection}{B.1 Recursive setup and Bellman
problem}

Let the recursive state vector be
\begin{equation}\protect\phantomsection\label{eq-b-state}{
 s_t = (w_{t-1},\sigma_{t-1}^w,a_t,z_t)^\top.
}\end{equation} I write the current control vector in primal form as \[
 u_t = (x_t,\pi_t^p,\pi_t^w,w_t,\sigma_t^w,\sigma_t^c,\sigma_t^n,\omega_t)^\top.
\] Here \(x_t\) is the output gap relative to the efficient
technology-driven reference allocation: \(x_t=y_t-y_t^f\). The primal
formulation treats the allocation variables as the policymaker's choice
objects. Once the optimal allocation is obtained, the nominal interest
rate is recovered from the TANK IS curve.

The time-invariant policy functions are \[
 x_t = X(s_t),\quad
 \pi_t^p = \Pi^p(s_t),\quad
 \pi_t^w = \Pi^w(s_t),\quad
 w_t = W(s_t),
\] \[
 \sigma_t^w = \Sigma^w(s_t),\quad
 \sigma_t^c = \Sigma^c(s_t),\quad
 \sigma_t^n = \Sigma^n(s_t),\quad
 \omega_t = \Omega(s_t).
\] Given the current endogenous states chosen by the policymaker, the
next-period state is \[
 s_{t+1} = (w_t,\sigma_t^w,a_{t+1},z_{t+1})^\top.
\] The continuation expectations are \[
 \hat{\Pi}^{p}_t \equiv E_t\left[\Pi^p(s_{t+1})\right],
 \qquad
 \hat{W}_t \equiv E_t\left[W(s_{t+1})\right],
 \qquad
 \hat{\Sigma}^{w}_t \equiv E_t\left[\Sigma^w(s_{t+1})\right].
\] The derivatives with respect to the endogenous next-state variables
are \[
\begin{aligned}
 \hat{\Pi}^{p}_{w,t} &\equiv \frac{\partial \hat{\Pi}^{p}_t}{\partial w_t},
&
 \hat{\Pi}^{p}_{\sigma,t} &\equiv \frac{\partial \hat{\Pi}^{p}_t}{\partial \sigma_t^w},\\
 \hat{W}_{w,t} &\equiv \frac{\partial \hat{W}_t}{\partial w_t},
&
 \hat{W}_{\sigma,t} &\equiv \frac{\partial \hat{W}_t}{\partial \sigma_t^w},\\
 \hat{\Sigma}^{w}_{w,t} &\equiv \frac{\partial \hat{\Sigma}^{w}_t}{\partial w_t},
&
 \hat{\Sigma}^{w}_{\sigma,t} &\equiv \frac{\partial \hat{\Sigma}^{w}_t}{\partial \sigma_t^w}.
\end{aligned}
\] These derivative terms are part of the maintained recursive problem.
They vanish only under additional approximations.

Under the own-lag wage-inflation identity, the wage-adjustment term
depends on \(\sigma_t^w - \sigma_{t-1}^w\).

The second-order period loss is
\begin{equation}\protect\phantomsection\label{eq-b-exact-loss}{
\begin{aligned}
\ell_t
 ={}&
 \frac{1}{2}
 \left\{
 (\gamma+\varphi)x_t^2 + \eta_p (\pi_t^p)^2 + \eta_w (\pi_t^w)^2
 \right\} \\
&+
 \frac{\lambda(1-\lambda)}{2}
 \left\{
 \gamma (\sigma_t^c)^2 + \left(\varphi+\frac{1}{\psi_w}\right) (\sigma_t^n)^2 + \eta_w (\sigma_t^w-\sigma_{t-1}^w)^2
 \right\}.
\end{aligned}
}\end{equation} The Bellman equation is
\begin{equation}\protect\phantomsection\label{eq-b-bellman}{
 V(s_t)
 =
 \min_{u_t}
 \left\{
 \ell_t + \beta E_t\left[V(s_{t+1})\right]
 \right\}
}\end{equation} subject to the canonical constraints
\begin{equation}\protect\phantomsection\label{eq-b-canonical-recursive}{
\begin{aligned}
 \pi_t^p &= \beta \hat{\Pi}^{p}_t + \kappa_p (w_t-a_t),\\
 \pi_t^w &= \pi_t^p + w_t - w_{t-1},\\
 w_t &= \frac{1}{\Theta_w}
 \left(
 w_{t-1} + \beta \hat{W}_t + \kappa_x x_t + \kappa_a a_t
 \right),\\
 \sigma_t^w &= \frac{1}{\Theta_\sigma}
 \left(
 \sigma_{t-1}^w + \beta \hat{\Sigma}^{w}_t + \frac{\gamma\kappa_w}{1-\lambda}\omega_t
 \right),\\
 \sigma_t^n &= -\psi_w \sigma_t^w,\\
 \sigma_t^c &= (1-\psi_w)\sigma_t^w + \frac{\omega_t}{1-\lambda},\\
 \omega_t &= a_t - w_t - z_t.
\end{aligned}
}\end{equation} The exogenous shocks evolve as \[
 a_t = \rho_a a_{t-1} + e_t^a,
 \qquad
 z_t = \rho_z z_{t-1} + e_t^z.
\] Because the nominal interest rate enters the canonical system only
through the IS relation, it can be recovered after solving the primal
problem. The implementability condition is
\begin{equation}\protect\phantomsection\label{eq-b-implementability}{
 i_t
 =
 E_t\pi_{t+1}^p+r_t^f + \gamma\left(E_t x_{t+1} - x_t\right)
 - \gamma\lambda\left(\sigma_t^c - E_t\sigma_{t+1}^c\right).
}\end{equation} Equivalently, \[
 i_t
 =
 E_t\pi_{t+1}^p+r_t^f + \gamma E_t\Delta x_{t+1} + \gamma\lambda E_t\Delta \sigma_{t+1}^c.
\]

\subsection*{B.2 Reduced recursive system and policy
instrument}\label{b.2-reduced-recursive-system-and-policy-instrument}
\addcontentsline{toc}{subsection}{B.2 Reduced recursive system and
policy instrument}

After eliminating static multipliers and applying the envelope
conditions, the exact discretionary problem reduces to two recursive
optimality equations. The first is the target condition for the
aggregate stabilization margin. The second is the recursion for the
distributional shadow price. The online appendix reports the
intermediate algebra. Let \(\mu_t^\sigma\) denote the multiplier on the
wage-dispersion constraint, written with its right-hand side minus
\(\sigma_t^w\) equal to zero.

Substituting the static multiplier eliminations and the envelope
conditions into the first-order condition for \(w_t\) gives the
operator-form target condition
\begin{equation}\protect\phantomsection\label{eq-b-target}{
 \begin{aligned}
 0 ={}&
 -\beta E_t\left[\eta_w\pi_{t+1}^w + \frac{\gamma+\varphi}{\kappa_x}x_{t+1}\right]
 + \left(\kappa_p + \beta\hat{\Pi}^{p}_{w,t}\right)\left(\eta_p\pi_t^p + \eta_w\pi_t^w\right)
 + \eta_w\pi_t^w \\
 &+ \left(1-\frac{\beta}{\Theta_w}\hat{W}_{w,t}\right)
 \frac{\Theta_w(\gamma+\varphi)}{\kappa_x}x_t 
 + \frac{\beta}{\Theta_\sigma}\hat{\Sigma}^{w}_{w,t}\mu_t^{\sigma}
 - \frac{\gamma\kappa_w}{(1-\lambda)\Theta_\sigma}\mu_t^{\sigma} 
 - \lambda\gamma(1-\psi_w)\sigma_t^w
 - \frac{\lambda\gamma}{1-\lambda}\omega_t.
 \end{aligned}
}\end{equation} This is the aggregate-wage target condition after the
static multiplier eliminations. Relative to a formulation that does not
put the cross-type wage-inflation term in the period loss, the displayed
aggregate-wage condition has the same operator form: the exact-loss
correction enters through the wage-dispersion shadow value
\(\mu_t^\sigma\) and its recursion below, not through a new direct term
in the first-order condition for \(w_t\).

Substituting the static eliminations and the envelope conditions into
the first-order condition for \(\sigma_t^w\) gives the recursive
equation for the distributional shadow price
\begin{equation}\protect\phantomsection\label{eq-b-mu-sigma}{
 \begin{aligned}
 \left(1-\frac{\beta}{\Theta_\sigma}\hat{\Sigma}^{w}_{\sigma,t}\right)\mu_t^{\sigma}
 ={}&
 \frac{\beta}{\Theta_\sigma}E_t\left[\mu_{t+1}^{\sigma}\right]
 + \beta\hat{\Pi}^{p}_{\sigma,t}\left(\eta_p\pi_t^p + \eta_w\pi_t^w\right)  - \beta\hat{W}_{\sigma,t}\frac{\gamma+\varphi}{\kappa_x}x_t \\
 &+ \lambda(1-\lambda)
 \left[(1+\beta)\eta_w + (\varphi\psi_w^2+\psi_w) + \gamma(1-\psi_w)^2\right]\sigma_t^w \\
 &- \lambda(1-\lambda)\eta_w\sigma_{t-1}^w
 - \lambda(1-\lambda)\beta\eta_w E_t\sigma_{t+1}^w + \lambda\gamma(1-\psi_w)\omega_t.
 \end{aligned}
}\end{equation} The second-order welfare-based objective generates the
lagged-state term proportional to \(\sigma_{t-1}^w\) and the lead term
proportional to \(E_t\sigma_{t+1}^w\) in this recursive equation. These
terms arise from the wage-adjustment penalty on changes in the wage gap.
The labor-dispersion term separately penalizes its level.

Together, these two equations form the reduced recursive optimality
system for the endogenous state block under discretion. A commitment
formulation would instead introduce lagged multipliers as additional
predetermined objects; that case is not solved here.

The primal problem delivers the optimal allocation. After solving for
the policy functions, the nominal interest rate is recovered from the
implementability condition \[
 i_t
 =
 E_t\pi_{t+1}^p+r_t^f + \gamma\left(E_t x_{t+1} - x_t\right)
 - \gamma\lambda\left(\sigma_t^c - E_t\sigma_{t+1}^c\right).
\] Here \(r_t^f\) is the natural interest-rate process associated with
natural output \(y_t^f\); in the baseline shock specification, it
depends on the technology process and not on the transfer shock. Since
\[
 E_t\sigma_{t+1}^c
 =
 E_t\left[(1-\psi_w)\sigma_{t+1}^w + \frac{\omega_{t+1}}{1-\lambda}\right],
\] monetary policy responds not only to expected aggregate activity but
also to the expected evolution of cross-type inequality.

\subsection*{B.3 Link to numerical
implementation}\label{b.3-link-to-numerical-implementation}
\addcontentsline{toc}{subsection}{B.3 Link to numerical implementation}

The reduced system above is the object solved in the passive-transfer
quantitative analysis. Recursive discretion is implemented with the
endogenous state vector \((w_{t-1},\sigma_{t-1}^w)\) after the static
multipliers are eliminated analytically. The online appendix reports the
full algebra and continuation-policy derivative terms in Part A, and the
numerical solution methods and diagnostics in Part B. Software
requirements and execution instructions are provided in the accompanying
replication documentation.

\section*{Appendix C. Proofs and Theoretical Benchmark
Derivations}\label{sec-appendix-c}
\addcontentsline{toc}{section}{Appendix C. Proofs and Theoretical
Benchmark Derivations}

This appendix proves the passive-transfer propositions and the
active-transfer results and reports the theoretical limits used in the
analysis. Throughout, assume that the recursive discretionary problem in
Appendix B is well defined, that the policy functions are differentiable
in a neighborhood of the deterministic steady state, and that the
interior first-order conditions derived in the online appendix hold.

For the proposition proofs, the arguments use the canonical recursive
constraints (\ref{eq-b-canonical-recursive}), the exact discretionary
target condition (\ref{eq-b-target}), and the shadow-value recursion for
wage dispersion (\ref{eq-b-mu-sigma}). These equations are not repeated
below.

\subsection*{C.1 Proof of Proposition
1}\label{c.1-proof-of-proposition-1}
\addcontentsline{toc}{subsection}{C.1 Proof of Proposition 1}

Fix the time-invariant continuation policy functions of the
discretionary equilibrium. The canonical constraints depend on lagged
endogenous variables through the aggregate real-wage state and the
cross-type wage-gap state. The shocks are first-order Markov. Hence
\(s_t=(w_{t-1},\sigma_{t-1}^w,a_t,z_t)^\top\) summarizes the inherited
objects needed to evaluate the current constraints and continuation
expectations.

To establish the feasible-set statement, take two initial states that
agree on \((w_{t-1},a_t,z_t)\) and differ only in the inherited gap,
denoted \(d\) and \(\widetilde d\). Suppose the same full current
allocation were feasible at both states. In particular, the allocation
would have the same \(w_t\), \(\sigma_t^w\), and \(\omega_t\). It would
therefore generate the same distribution of next-period states and the
same continuation expectations. Subtracting the two wage-dispersion
constraints gives \[
0=\frac{d-\widetilde d}{\Theta_\sigma},
\] which is impossible for \(d\ne\widetilde d\). Thus the full current
allocation cannot be held fixed as the inherited gap changes,
conditional on the same continuation functions. This argument concerns
implementability and does not impose a nonzero response on any
particular allocation component or instrument.

For the welfare statement, use \(\sigma_t^n=-\psi_w\sigma_t^w\) in the
period loss. Every term in that loss is nonnegative under the maintained
parameter restrictions. Retaining only labor dispersion and cross-type
wage-inflation dispersion gives \[
\ell_t
\ge
\frac{\lambda(1-\lambda)}{2}
\left\{
W_n(\sigma_t^w)^2
+\eta_w(\sigma_t^w-d)^2
\right\}.
\] Completing the square yields
\begin{equation}\protect\phantomsection\label{eq-passive-legacy-completion}{
\begin{aligned}
W_n(\sigma_t^w)^2+\eta_w(\sigma_t^w-d)^2
={}&(W_n+\eta_w)
\left(\sigma_t^w-\frac{\eta_w}{W_n+\eta_w}d\right)^2\\
&+\frac{W_n\eta_w}{W_n+\eta_w}d^2.
\end{aligned}
}\end{equation} The lower bound holds for every current allocation, so
it also holds for the discretionary allocation. Subsequent period losses
are nonnegative. Along the path with no further shocks from
\(s=d e_\sigma\), their discounted sum therefore satisfies
(\ref{eq-passive-legacy-value-bound}). The homogeneous linear
equilibrium remains at the symmetric steady state when \(s=0\) and no
innovations occur, giving \(V^0(0)=0\). Since both \(W_n\) and
\(\eta_w\) are positive, the bound is strictly positive for a nonzero
inherited gap.

The result can equivalently be expressed in terms of value curvature.
Write the period loss along the linear equilibrium as
\(\ell_t=\tfrac12s_t^\top Qs_t\), where \(Q\) is positive semidefinite.
Finite discounted loss implies \[
V^0(s)=\frac12s^\top\mathcal P s,
\qquad
\mathcal P=\sum_{j=0}^{\infty}\beta^j(A^j)^\top Q A^j.
\] It follows that
\begin{equation}\protect\phantomsection\label{eq-passive-value-curvature}{
e_\sigma^\top\mathcal P e_\sigma
\ge
\lambda(1-\lambda)
\frac{W_n\eta_w}{W_n+\eta_w}>0.
}\end{equation} With additive, zero-mean innovations whose future
distribution is independent of the current state, the associated
stochastic equilibrium value is \(V(s)=\tfrac12s^\top\mathcal P s+c\),
where \(c\) is the contribution of future innovations. The curvature is
unchanged. The first derivative of \(V\) with respect to the inherited
gap is zero at the symmetric origin, as it is for any quadratic loss
centered at that origin; positive curvature, rather than a nonzero
derivative at the origin, establishes payoff relevance. QED.

\subsection*{C.2 Proof of Proposition
2}\label{c.2-proof-of-proposition-2}
\addcontentsline{toc}{subsection}{C.2 Proof of Proposition 2}

The nominal-rate row in (\ref{eq-passive-nominal-rate-row}) follows from
the Fisher relation and the linear continuation expectation: \[
i_t=r_t+E_t\pi_{t+1}^p
=R s_t+P A s_t.
\] Now choose a vector \(v\) satisfying
(\ref{eq-passive-policy-rank-conditions}). Since the linear equilibrium
is defined on a neighborhood of admissible initial states, \(s\) and
\(\widetilde s=s+\varepsilon v\) are both admissible for any interior
\(s\) and sufficiently small nonzero \(\varepsilon\). Their current
aggregate outcomes satisfy \[
H\widetilde s-Hs=\varepsilon Hv=0.
\] Their inherited wage gaps and nominal rates instead differ by \[
e_\sigma^\top(\widetilde s-s)=\varepsilon e_\sigma^\top v\ne0,
\qquad
I\widetilde s-Is=\varepsilon Iv\ne0.
\] A function of the common aggregate outcome \(Hs=H\widetilde s\) must
assign the same rate to both states. It therefore cannot coincide with
the implementing nominal-rate function at both states. This proves the
proposition. QED.

The conditions also admit a useful rank formulation. Since \(H\) has
rank three, its nullspace is one-dimensional. For any nonzero vector
\(v\) in that nullspace, \[
Iv\ne0
\quad\Longleftrightarrow\quad
\operatorname{rank}\begin{bmatrix}H\\I\end{bmatrix}=4,
\] and \[
e_\sigma^\top v\ne0
\quad\Longleftrightarrow\quad
\operatorname{rank}\begin{bmatrix}H\\e_\sigma^\top\end{bmatrix}=4.
\] The first condition rules out recovering the rate from the three
aggregate outcomes alone. The second ensures that the indistinguishable
states differ in the inherited wage gap and that adding this gap
recovers all four state coordinates. These are separate restrictions.
Neither is implied solely by a nonzero direct coefficient
\(I e_\sigma\).

The proposition compares admissible initial states of the equilibrium
system. A stationary projection instead compares states generated by the
specified stochastic processes. If those processes explore only a proper
subspace of the four-dimensional state space, the relevant spanning test
must be restricted to that subspace. A positive-definite stationary
state covariance allows the full-state rank test to apply to the
stationary projection as well. Online Appendix B.3 reports the benchmark
rank checks and the stationary projection exercise separately.

\subsection*{C.3 Representative-agent
limit}\label{c.3-representative-agent-limit}
\addcontentsline{toc}{subsection}{C.3 Representative-agent limit}

Set \(\lambda = 0\). Then the IS equation collapses to \[
x_t = E_t x_{t+1} - \frac{1}{\gamma}(r_t-r_t^f),
\] and the loss function reduces to \[
\mathcal L
=
E_0\sum_{t=0}^{\infty}\beta^t
\frac{1}{2}
\left\{
(\gamma+\varphi)x_t^2 + \eta_p(\pi_t^p)^2 + \eta_w(\pi_t^w)^2
\right\}.
\] The distributional variables no longer affect aggregate dynamics or
welfare. The policy problem becomes the standard dual-rigidity
representative-agent stabilization problem.

\subsection*{C.4 Flexible-wage and sticky-price-only TANK
limits}\label{c.4-flexible-wage-and-sticky-price-only-tank-limits}
\addcontentsline{toc}{subsection}{C.4 Flexible-wage and
sticky-price-only TANK limits}

There are two related limits. The first is the joint limit
\(\eta_w\downarrow0\) and \(\psi_w\to\infty\), which removes
wage-adjustment resource costs and makes labor services perfect
substitutes. Then cross-type wage dispersion vanishes,
\(\sigma_t^w \to 0\), but the static dispersion mapping implies \[
\sigma_t^c \to \frac{\varphi}{\varphi+\gamma}\frac{\omega_t}{1-\lambda}.
\] Hence consumption dispersion generally survives unless
\(\omega_t = 0\). The flexible-wage benchmark is generally not
equivalent to a representative-agent model.

Because the canonical coefficients use \(\kappa_w=\psi_w/\eta_w\), the
sticky-price-only TANK benchmark should be read as the limit
\(\eta_w\downarrow 0\) with finite \(\psi_w\), so \(\kappa_w\to\infty\),
rather than as a literal finite-coefficient parameterization with
\(\eta_w=0\). In this limiting system, the wage-dispersion block becomes
static rather than inertial: \[
\sigma_t^w
=
\frac{\gamma}{1-\gamma + \psi_w(\varphi+\gamma)}
\frac{\omega_t}{1-\lambda}.
\] Thus wage dispersion no longer carries an inherited state, but
consumption dispersion still depends on the contemporaneous net wedge.
The sticky-price-only TANK benchmark is therefore analytically distinct
from the representative-agent benchmark.

\subsection*{C.5 Active-transfer separation and proof of Proposition
3}\label{c.5-active-transfer-separation-and-proof-of-proposition-3}
\addcontentsline{toc}{subsection}{C.5 Active-transfer separation and
proof of Proposition 3}

With active transfers, the policy authority chooses the aggregate
allocation and the active distributional wedge. The recursive state can
be written as \((w_{t-1},\sigma_{t-1}^w,a_t,z_t)^\top\), but the active
wedge \[
\omega_t^A=a_t-w_t-z_t-f_t
\] lets the policy authority choose the distributional wedge
independently of \((a_t,w_t,z_t)\) and then recover \(f_t\).

The loss separates as \[
\ell_t=\ell_t^{agg}+\ell_t^{dist},
\] with \(\ell_t^{agg}\) depending only on \((x_t,\pi_t^p,\pi_t^w)\) and
\(\ell_t^{dist}\) depending only on
\((\sigma_t^c,\sigma_t^n,\sigma_t^w,\sigma_{t-1}^w)\). The constraints
also split. The aggregate block is \[
\pi_t^p = \beta E_t \pi_{t+1}^p + \kappa_p(w_t-a_t),
\] \[
w_t = \frac{1}{\Theta_w}\left(w_{t-1}+\beta E_t w_{t+1}+\kappa_x x_t + \kappa_a a_t\right),
\] and the IS equation recovers the implementing rate. The distribution
block is \[
\sigma_t^w
=
\frac{1}{\Theta_\sigma}
\left[
\sigma_{t-1}^w+\beta E_t\sigma_{t+1}^w
+\frac{\gamma\kappa_w}{1-\lambda}\omega_t^A
\right],
\] \[
\sigma_t^c=(1-\psi_w)\sigma_t^w+
\frac{\omega_t^A}{1-\lambda},
\qquad
\sigma_t^n=-\psi_w\sigma_t^w.
\] No constraint links the choice of \((x_t,\pi_t^p,\pi_t^w,w_t)\) to
\((\sigma_t^w,\sigma_t^c,\sigma_t^n,\omega_t^A)\) once \(f_t\) is
unrestricted.

The Bellman problem is therefore additively separable. Up to
policy-independent terms, \[
\begin{aligned}
V^A(w_{t-1},\sigma_{t-1}^w,a_t,z_t)
=&
\min_{\{x_t,\pi_t^p,\pi_t^w,w_t\}}
\left\{\ell_t^{agg}+\beta E_t V^{agg}(w_t,a_{t+1})\right\}\\
&+
\min_{\{\sigma_t^w,\sigma_t^c,\sigma_t^n,\omega_t^A\}}
\left\{\ell_t^{dist}+\beta E_t V^{dist}(\sigma_t^w)\right\}.
\end{aligned}
\] This proves the value-function decomposition in Proposition 3. The
aggregate minimization is the dual-rigidity RANK discretionary
stabilization problem because the cross-type dispersion variables do not
enter its constraints or objective. The shock \(z_t\) is absent from the
allocation problem but remains in \[
f_t^*=a_t-w_t^*-z_t-\omega_t^{A,*}.
\] The implementing nominal rate is \[
i_t^*=E_t\pi_{t+1}^{p,*}+r_t^f+\gamma(E_tx_{t+1}^*-x_t^*)
-\gamma\lambda(\sigma_t^{c,*}-E_t\sigma_{t+1}^{c,*}).
\] Thus allocation separation coexists with instrument coupling.

\subsection*{C.6 Symmetric state, RI, and proof of Proposition
4}\label{c.6-symmetric-state-ri-and-proof-of-proposition-4}
\addcontentsline{toc}{subsection}{C.6 Symmetric state, RI, and proof of
Proposition 4}

If \(\sigma_{t-1}^w=0\), setting \(\sigma_t^w=E_t\sigma_{t+1}^w=0\) and
\(\omega_t^A=0\) satisfies the wage-dispersion equation. The static
dispersion equations then give \(\sigma_t^c=\sigma_t^n=0\). This proves
Corollary 1.

Now suppose \(\sigma_{-1}^w\neq0\) and impose the RI path \[
\sigma_t^w=0\quad \text{for all }t\ge0.
\] At \(t=0\), the wage-dispersion equation implies \[
0
=
\frac{1}{\Theta_\sigma}
\left[
\sigma_{-1}^w+\frac{\gamma\kappa_w}{1-\lambda}\omega_0^A
\right],
\] so \[
\omega_0^A=-\frac{1-\lambda}{\gamma\kappa_w}\sigma_{-1}^w.
\] The impact consumption gap is therefore \[
\sigma_0^c
=
\frac{\omega_0^A}{1-\lambda}
=-\frac{1}{\gamma\kappa_w}\sigma_{-1}^w.
\] Thus RI implements the RANK distributional allocation only after the
impact period.

For Proposition 4, eliminate the active wedge from the wage-dispersion
equation: \[
\omega_t^A
=
\frac{1-\lambda}{\gamma\kappa_w}
\left(
\Theta_\sigma\sigma_t^w-\sigma_{t-1}^w-\beta E_t\sigma_{t+1}^w
\right).
\] Substitution into the static consumption-dispersion equation gives \[
\sigma_t^c
=A_\sigma\sigma_t^w
-B_\sigma\sigma_{t-1}^w
-C_\sigma E_t\sigma_{t+1}^w,
\] where \[
A_\sigma
=1-\psi_w+\frac{\Theta_\sigma}{\gamma\kappa_w},
\quad
B_\sigma=\frac{1}{\gamma\kappa_w},
\quad
C_\sigma=\frac{\beta}{\gamma\kappa_w}.
\] Using the definition of \(\Theta_\sigma\), \[
A_\sigma
=\frac{1+\beta}{\gamma\kappa_w}
+\frac{1+\psi_w\varphi}{\gamma}>0.
\]

For a deterministic distributional transition, define the unscaled loss
\(\widetilde{\mathcal L}=2\sum_{t=0}^{\infty}\beta^t\ell_t^{dist}/[\lambda(1-\lambda)]\).
Starting at RI, perturb only the impact wage gap: set
\(\sigma_0^w=\varepsilon\) and keep \(\sigma_t^w=0\) for \(t\ge1\). At
\(\varepsilon=0\), the derivative of the unscaled distributional loss is
\[
\frac{d\widetilde{\mathcal L}}{d\varepsilon}\bigg|_{\varepsilon=0}
=
-2\sigma_{-1}^w
\left(\gamma A_\sigma B_\sigma+\eta_w\right).
\] The term in parentheses is positive because \(A_\sigma>0\),
\(B_\sigma>0\), and \(\eta_w>0\). Choosing \(\varepsilon\) with the same
sign as \(\sigma_{-1}^w\) lowers loss. Labor dispersion and future
adjustment costs are second order at this perturbation. RI is therefore
not a local optimum from a nonzero legacy state.

\subsection*{C.7 Proof of Proposition
5}\label{c.7-proof-of-proposition-5}
\addcontentsline{toc}{subsection}{C.7 Proof of Proposition 5}

Write \(s_t=\sigma_t^w\) and \(W_n=\varphi\psi_w^2+\psi_w\). After
eliminating the active wedge, the unscaled distributional loss is \[
\widetilde{\mathcal L}
=
\sum_{t=0}^{\infty}\beta^t
\left[
\gamma(\sigma_t^c)^2
+W_n s_t^2
+\eta_w(s_t-s_{t-1})^2
\right],
\] where \[
\sigma_t^c
=
A_\sigma s_t-B_\sigma s_{t-1}-C_\sigma s_{t+1}.
\] The CES-consistent coefficient and \(\kappa_w=\psi_w/\eta_w\) imply
\[
\Theta_\sigma-\gamma\kappa_w(\psi_w-1)
=1+\beta+\frac{W_n}{\eta_w}
\equiv \mathcal K.
\] It follows that \[
\sigma_t^c
=
\frac{\mathcal K s_t-s_{t-1}-\beta s_{t+1}}
{\gamma\kappa_w}.
\]

Dropping the nonnegative consumption-dispersion term gives a lower bound
on \(\widetilde{\mathcal L}\). The Euler equation for the remaining
strictly convex smoothing problem is \[
\mathcal K s_t=s_{t-1}+\beta s_{t+1}.
\] Its bounded solution is \(s_t=F^*s_{t-1}\), where \[
F^*
=\frac{\mathcal K-\sqrt{\mathcal K^2-4\beta}}{2\beta}
=\frac{2}{\mathcal K+\sqrt{\mathcal K^2-4\beta}}.
\] This path also sets \(\sigma_t^c=0\), so it attains the lower bound
and is the global optimum of the full distributional problem. Because
the same stationary continuation is optimal after every inherited state,
it is both the date-0 commitment solution and the Markov-perfect
discretionary solution. Substitution into the value recursion gives \[
P^*=\eta_w(1-F^*).
\] Finally, the implementing active wedge is \[
\omega_t^{A,*}=(1-\lambda)(\psi_w-1)s_t,
\] and the consumption-dispersion contribution to the implementing
interest rate is zero. This proves Proposition 5.


\begin{CSLReferences}{1}{1}
\bibitem[\citeproctext]{ref-OptimalMonetaryPolicy-Acharya-2023}
Acharya, Sushant, Édouard Challe, and Keshav Dogra. 2023. {``Optimal
Monetary Policy According to {HANK}.''} \emph{American Economic Review}
113 (7): 1741--82. \url{https://doi.org/10.1257/aer.20200239}.

\bibitem[\citeproctext]{ref-LimitedAssetMarketParticipation-Ascari-2017}
Ascari, Guido, Andrea Colciago, and Lorenza Rossi. 2017. {``Limited
Asset Market Participation, Sticky Wages, and Monetary Policy.''}
\emph{Economic Inquiry} 55 (2): 878--97.
\url{https://doi.org/10.1111/ecin.12424}.

\bibitem[\citeproctext]{ref-Auclert2019}
Auclert, Adrien. 2019. {``Monetary Policy and the Redistribution
Channel.''} \emph{American Economic Review} 109 (6): 2333--67.
\url{https://doi.org/10.1257/aer.20160137}.

\bibitem[\citeproctext]{ref-FiscalMonetaryPolicy-Auclert-2025}
Auclert, Adrien, Matthew Rognlie, and Ludwig Straub. 2025. {``Fiscal and
{Monetary} {Policy} with {Heterogeneous} {Agents}.''} \emph{Annual
Review of Economics} 17 (Volume 17, 2025): 539--62.
\url{https://doi.org/10.1146/annurev-economics-091624-044646}.

\bibitem[\citeproctext]{ref-InequalityBusinessCycles-Bhandari-2021}
Bhandari, Anmol, David Evans, Mikhail Golosov, and Thomas J. Sargent.
2021. {``Inequality, Business Cycles, and Monetary-Fiscal Policy.''}
\emph{Econometrica} 89 (6): 2559--99.
\url{https://doi.org/10.3982/ECTA16414}.

\bibitem[\citeproctext]{ref-Bilbiie2008}
Bilbiie, Florin O. 2008. {``Limited Asset Markets Participation,
Monetary Policy and (Inverted) Aggregate Demand Logic.''} \emph{Journal
of Economic Theory} 140 (1): 162--96.
\url{https://doi.org/10.1016/j.jet.2007.07.008}.

\bibitem[\citeproctext]{ref-MonetaryPolicyHeterogeneity-Bilbiie-2025}
Bilbiie, Florin O. 2025. {``Monetary Policy and Heterogeneity: An
Analytical Framework.''} \emph{Review of Economic Studies} 92 (4):
2398--436. \url{https://doi.org/10.1093/restud/rdae066}.

\bibitem[\citeproctext]{ref-StabilizationVsRedistribution-Bilbiie-2024}
Bilbiie, Florin O., Tommaso Monacelli, and Roberto Perotti. 2024.
{``Stabilization Vs. {Redistribution}: {The} Optimal Monetary--Fiscal
Mix.''} \emph{Journal of Monetary Economics}, Monetary {Policy}
challenges for {European} {Macroeconomies}, vol. 147 (October): 103623.
\url{https://doi.org/10.1016/j.jmoneco.2024.103623}.

\bibitem[\citeproctext]{ref-OptimalMonetaryPolicy-Bilbiie-2021}
Bilbiie, Florin O., and Xavier Ragot. 2021. {``Optimal Monetary Policy
and Liquidity with Heterogeneous Households.''} \emph{Review of Economic
Dynamics}, Special {Issue} in {Memory} of {Alejandro} {Justiniano}, vol.
41 (July): 71--95. \url{https://doi.org/10.1016/j.red.2020.10.003}.

\bibitem[\citeproctext]{ref-RealWageRigidities-Blanchard-2007}
Blanchard, Olivier, and Jordi Galí. 2007. {``Real Wage Rigidities and
the New Keynesian Model.''} \emph{Journal of Money, Credit and Banking}
39 (s1): 35--65. \url{https://doi.org/10.1111/j.1538-4616.2007.00015.x}.

\bibitem[\citeproctext]{ref-NewKeynesianTransmission-Broer-2020}
Broer, Tobias, Niels-Jakob Harbo Hansen, Per Krusell, and Erik Öberg.
2020. {``The New Keynesian Transmission Mechanism: A Heterogeneous-Agent
Perspective.''} \emph{Review of Economic Studies} 87 (1): 77--101.
\url{https://doi.org/10.1093/restud/rdy060}.

\bibitem[\citeproctext]{ref-Colciago2011}
Colciago, Andrea. 2011. {``Rule-of-Thumb Consumers Meet Sticky Wages.''}
\emph{Journal of Money, Credit and Banking} 43 (2-3): 325--53.
\url{https://doi.org/10.1111/j.1538-4616.2011.00376.x}.

\bibitem[\citeproctext]{ref-OptimalMonetaryPolicy-Davila-2023}
Dávila, Eduardo, and Andreas Schaab. 2023. \emph{Optimal {Monetary}
{Policy} with {Heterogeneous} {Agents}: {Discretion}, {Commitment}, and
{Timeless} {Policy}}. No. w30961. National Bureau of Economic Research.
\url{https://doi.org/10.3386/w30961}.

\bibitem[\citeproctext]{ref-HeterogeneityAggregateFluctuations-Debortoli-2024}
Debortoli, Davide, and Jordi Galí. 2025. {``Heterogeneity and
{Aggregate} {Fluctuations}: {Insights} from {TANK} {Models}.''}
\emph{NBER Macroeconomics Annual} 39 (1): 307--57.
\url{https://doi.org/10.1086/735272}.

\bibitem[\citeproctext]{ref-OptimalMonetaryPolicy-Erceg-2000}
Erceg, Christopher J., Dale W. Henderson, and Andrew T. Levin. 2000.
{``Optimal Monetary Policy with Staggered Wage and Price Contracts.''}
\emph{Journal of Monetary Economics} 46 (2): 281--313.
\url{https://doi.org/10.1016/S0304-3932(00)00028-3}.

\bibitem[\citeproctext]{ref-GerkeGiesenLozejRoettger2024}
Gerke, Rafael, Sebastian Giesen, Matija Lozej, and Joost Röttger. 2024.
\emph{On Household Labour Supply in Sticky-Wage HANK Models}. Discussion
Paper 01/2024. Deutsche Bundesbank.
\url{https://doi.org/10.2139/ssrn.4744547}.

\bibitem[\citeproctext]{ref-IdaOkano2024}
Ida, Daisuke, and Mitsuhiro Okano. 2024. {``Does Nominal Wage Stickiness
Affect Fiscal Multiplier in a Two-Agent New Keynesian Model?''}
\emph{The B.E. Journal of Macroeconomics} 24 (2): 883--928.
\url{https://doi.org/10.1515/bejm-2023-0213}.

\bibitem[\citeproctext]{ref-KaplanMollViolante2018}
Kaplan, Greg, Benjamin Moll, and Giovanni L. Violante. 2018. {``Monetary
Policy According to HANK.''} \emph{American Economic Review} 108 (3):
697--743. \url{https://doi.org/10.1257/aer.20160042}.

\bibitem[\citeproctext]{ref-LaOMorrison2025}
La'O, Jennifer, and Wendy Morrison. 2025. {``Optimal Monetary Policy
with Redistribution.''} May 5.
\url{https://www.wendyamorrison.com/_files/ugd/041ab5_4d70adf52ed54f4a932cb1ca9459754c.pdf}.

\bibitem[\citeproctext]{ref-MatsuiYoshimi2013}
Matsui, Muneya, and Taiyo Yoshimi. 2013. {``Heterogeneity in Wage
Rigidity and Monetary Policy.''} \emph{Review of Integrative Business
and Economics Research} 2 (2): 491--520.
\url{https://buscompress.com/uploads/3/4/9/8/34980536/riber_b13-188__491-520_.pdf}.

\bibitem[\citeproctext]{ref-MatsuiYoshimi2015}
Matsui, Muneya, and Taiyo Yoshimi. 2015. {``Macroeconomic Dynamics in a
Model with Heterogeneous Wage Contracts.''} \emph{Economic Modelling}
49: 72--80. \url{https://doi.org/10.1016/j.econmod.2015.03.014}.

\bibitem[\citeproctext]{ref-McKayWolfJEP2023}
McKay, Alisdair, and Christian K. Wolf. 2023. {``Monetary Policy and
Inequality.''} \emph{Journal of Economic Perspectives} 37 (1): 121--44.
\url{https://doi.org/10.1257/jep.37.1.121}.

\bibitem[\citeproctext]{ref-OptimalPolicyRules-McKay-a}
McKay, Alisdair, and Christian K. Wolf. 2026. {``Optimal {Policy}
{Rules} in {HANK}.''} April 10.
\url{https://alisdairmckay.com/pdfs/hank_optpol.pdf}.

\bibitem[\citeproctext]{ref-Miyazaki2026}
Miyazaki, Kenji. 2026. {``When Redistribution Becomes a State Variable:
Monetary-Fiscal Stabilization with Type-Specific Sticky Wages.''}
\url{https://arxiv.org/abs/2605.15614v1}.

\bibitem[\citeproctext]{ref-OptimalMonetaryPolicy-Nistico-2016}
Nisticò, Salvatore. 2016. {``Optimal {Monetary} {Policy} and {Financial}
{Stability} in a {Non}-{Ricardian} {Economy}.''} \emph{Journal of the
European Economic Association} 14 (5): 1225--52.
\url{https://doi.org/10.1111/jeea.12182}.

\end{CSLReferences}
\end{document}